\documentclass[journal]{IEEEtran}

\IEEEoverridecommandlockouts                              

\usepackage{color}

\title{
Hierarchical Multi-Agent MCTS for Safety-Critical Coordination in Mixed-Autonomy Roundabouts
}

\author{Zhihao Lin$^{1}$, Jianglin Lan$^{1,\S}$, Shuo Liu$^{2}$,~\IEEEmembership{Student Member,~IEEE}, Zhen Tian$^{1}$, Dezong Zhao$^{1}$,~\IEEEmembership{Senior Member,~IEEE}, and Chongfeng~Wei$^{1}$,~\IEEEmembership{Senior Member,~IEEE}
\thanks{$^{1}$Zhihao Lin, Jianglin Lan, Zhen Tian, Dezong Zhao, and Chongfeng Wei are with James Watt School of Engineering, University of Glasgow, Glasgow G12 8QQ, United Kingdom.}%
\thanks{$^{2}$Shuo Liu is with Boston University, Brookline, MA, USA.}

\thanks{$^{\S}$Corresponding author. Jianglin Lan (e-mail: Jianglin.Lan@glasgow.ac.uk)}%
}

\definecolor{lightcyan}{RGB}{216,241,227}
\newcommand{\spm}[2]{#1\,{\scriptsize$\pm$\,#2}}
\usepackage{arydshln}
\usepackage{booktabs}
\usepackage{colortbl}  
\usepackage{bm}

\usepackage{array} 
\usepackage{makecell} 
\usepackage{algorithm}  
\usepackage{algpseudocode}
\usepackage{multirow}
\usepackage{epsfig} 
\usepackage{amsmath} 
\DeclareMathOperator*{\argmax}{arg\,max}
\usepackage{amsfonts} 

\usepackage[flushleft]{threeparttable}
\usepackage{graphicx}
\usepackage{hyperref}
\usepackage{tabularx} 
\usepackage{cite}
\usepackage{subcaption}
\begin{document}

\maketitle

\begin{abstract}
Navigating unsignalized roundabouts in mixed-autonomy traffic presents significant challenges due to dense vehicle interactions, lane-changing complexities, and behavioral uncertainties of human-driven vehicles (HDVs). This paper proposes a safety-critical decision-making framework for connected and automated vehicles (CAVs) navigating dual-lane roundabouts alongside HDVs. We formulate the problem as a multi-agent Markov Decision Process and develop a hierarchical safety assessment mechanism that evaluates three critical interaction types: CAV-to-CAV (C2C), CAV-to-HDV (C2H), and CAV-to-Boundary (C2B). A key contribution is our lane-specific uncertainty model for HDVs, which captures distinct behavioral patterns between inner and outer lanes, with outer-lane vehicles exhibiting 2.3× higher uncertainty due to less constrained movements. We integrate this safety framework with a multi-agent Monte Carlo Tree Search (MCTS) algorithm that employs safety-aware pruning to eliminate high-risk trajectories while maintaining computational efficiency. The reward function incorporates Shapley value-based credit assignment to balance individual performance with group coordination.  Extensive simulation results validate the effectiveness of the proposed approach under both fully autonomous (100\% AVs) and mixed traffic (50\% AVs + 50\% HDVs) conditions. Compared to benchmark methods, our framework consistently reduces trajectory deviations across all AVs and significantly lowers the rate of Post-Encroachment Time (PET) violations, achieving only 1.0\% in the fully autonomous scenario and 3.2\% in the mixed traffic setting.
\end{abstract}


\section{INTRODUCTION}

\IEEEPARstart{N}{avigation} at roundabouts presents unique challenges for Connected and Autonomous Vehicles (CAVs)~\cite{alkheder2020driver}, particularly in mixed traffic where both CAVs and human-driven vehicles (HDVs) must safely coordinate through complex circular geometries and multiple merging points~\cite{10394462,galceran2015multipolicy}. The difficulty lies in managing several critical interactions simultaneously while ensuring safety and efficiency in a highly dynamic setting with continuous merging, lane-changing, and exit decisions~\cite{polders2015identifying}. This challenge is further amplified by intricate interaction patterns induced by roundabout-specific geometric constraints and yielding rules. Therefore, a comprehensive understanding of both deterministic CAV behaviors and uncertain HDV behaviors is essential~\cite{BADUE2021113816,inproceedings}.

Traditional approaches to roundabout management often rely on rule-based decision-making methods, which generate conflict-free navigation sequences through preset yielding regulations and lane assignment strategies~\cite{choi2024calibration,shi2025safety}. These methods struggle to capture the complex decision-making behaviors of human drivers, where yielding involves significant behavioral variability. Conventional priority-based strategies~\cite{azimi2013v2v} ensure safety by enforcing strict yielding rules, but may substantially increase the requirement for onboard communication quality. More sophisticated rule-based approaches incorporating machine learning and static- and dynamic-constraint-based optimization have been proposed, yet their effectiveness has only been verified in simple single-lane roundabouts~\cite{10589570}.

The application of machine learning to roundabout navigation has advanced significantly, yet critical gaps remain between theory and practical deployment~\cite{11051098,11008885}. While deep reinforcement learning methods, including multi-agent deep deterministic policy gradient approaches, have shown success in controlled simulations~\cite{9693175,11120451,cai2022dq}, their performance degrades when faced with the geometric diversity and behavioral uncertainty of real-world roundabouts~\cite{10607945}. The core challenge lies not in learning capability itself, but in the mismatch between underlying assumptions and the requirements of safety-critical traffic coordination~\cite{ferrarotti2024autonomous,9311168,9565031}. Specifically, neural network-based approaches exhibit three key limitations~\cite{tian2024balanced}. First, the exponential growth of the state-action space with vehicle density makes it computationally prohibitive to cover all safety-critical scenarios during training~\cite{11051098,11008885}. Second, limited interpretability prevents real-time safety verification, a non-negotiable requirement in mixed autonomy~\cite{9693175,cai2022dq}. Third, the sim-to-real gap is severe in roundabouts due to the interplay between lane-changing decisions and continuous control, leading to unexpected behaviors when policies encounter novel traffic patterns.

\begin{figure*}[t]
      \centering
      \includegraphics[width=0.9\linewidth]{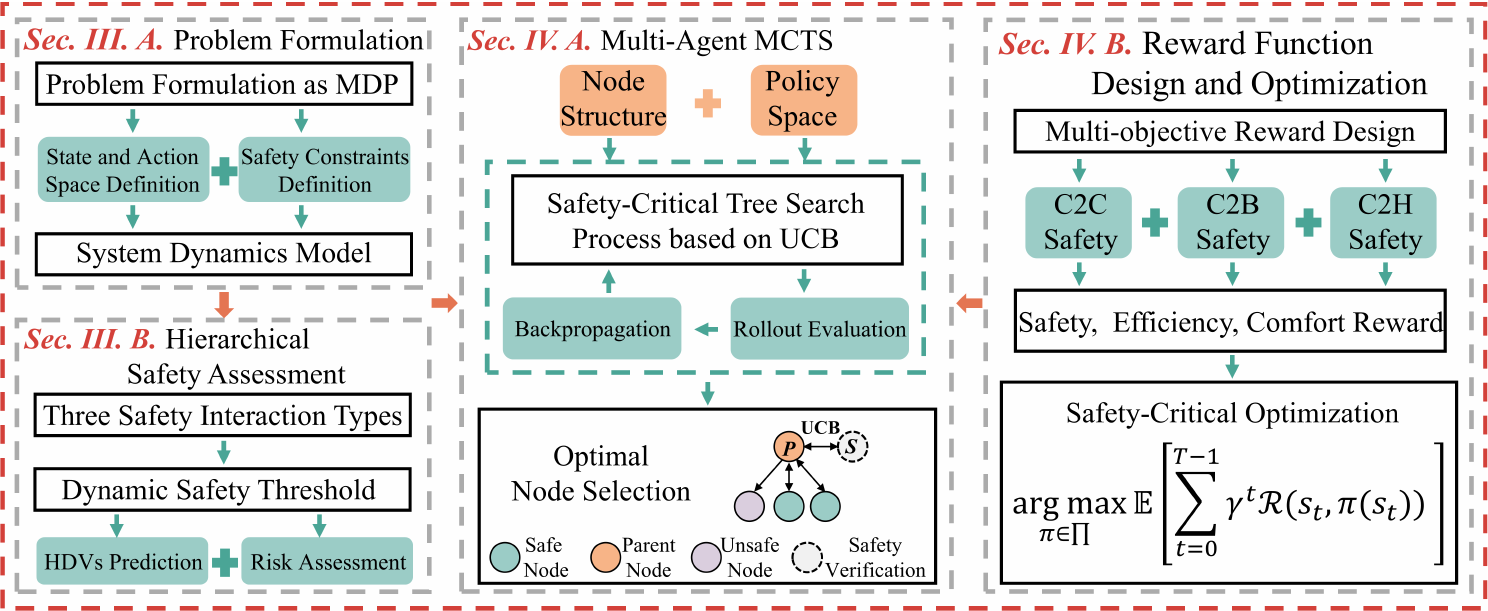}
      \vspace{-0.1cm}
        \caption{Overview of the proposed safety-critical decision making framework based on MCTS for roundabout navigation.}
      \label{fig2_toc}
\end{figure*}

In complex environments like roundabouts, vehicle interactions are often modeled using game-theoretic reasoning, aiming to capture both cooperative and competitive behaviors among drivers~\cite{zhu2023bi, 9557770, 9492817, ding2019multivehicle}. Frameworks such as Nash and Stackelberg games~\cite{pourabdollah2017calibration} enable multi-agent planning under strategic assumptions. However, real-world scenarios introduce significant deviations from these assumptions—human drivers may act irrationally, interpret gaps inconsistently, or delay exit decisions~\cite{Bie2024}, leading to mismatches between model predictions and actual behavior. Furthermore, the non-uniqueness of game-theoretic solutions~\cite{9781400} can hinder reliable coordination, particularly in multi-lane roundabouts where rapid decision-making is crucial for maintaining safety and efficiency~\cite{9456983}. Beyond learning and game-theoretic approaches, safety-critical control methods have also been widely investigated to provide formal safety guarantees. Representative examples include Control Barrier Functions (CBFs) \cite{liu2023iterative, liu2024feasibility}, reachability analysis \cite{bansal2017hamilton}, and formal methods. In contrast, our work addresses safety through a Monte Carlo Tree Search (MCTS) framework with hierarchical risk assessment, tailored to mixed traffic at unsignalized roundabouts.

MCTS~\cite{browne2012survey} has emerged as a promising approach by marrying the learning-based and game-theoretic methods for interactive navigation~\cite{lenz2016tactical}. Unlike traditional DRL which requires extensive offline training, MCTS can efficiently explore the action space through online planning, making it particularly suitable for the dynamic and geometry-dependent nature of roundabout environments~\cite{10186066,9737041}. The algorithm's inherent ability to balance exploration and exploitation makes it suitable for handling the uncertainties in mixed roundabout traffic, where vehicles are required to continuously make decisions about lane positioning, gap acceptance, and exit timing~\cite{hayes2021risk,11130528}. However, current MCTS implementations such as~\cite{weingertner2020monte,wang2020speed} often fall short in addressing comprehensive safety considerations and face significant scalability challenges in multi-agent scenarios involving multiple lanes and exit options.

This paper introduces a safety-critical multi-agent MCTS framework
for coordinating mixed traffic at dual-lane roundabouts. The framework
addresses the unique challenges posed by roundabout geometry, including
lane-specific uncertainty modeling, exit proximity effects, and
complex interactions between inner and outer lane vehicles.
Unlike prior MCTS-based planners that apply uniform
safety checks or treat all traffic participants identically, our
approach introduces a \emph{structurally differentiated} design:
safety assessment, uncertainty modeling, and reward allocation are
tailored to the specific interaction type (CAV-to-CAV (C2C)~\cite{10244038}, CAV-to-HDV 
(C2H)~\cite{gao2024trajectory} and CAV-to-Boundary 
(C2B)~\cite{10168227}) and
lane context, enabling these components to reinforce each other
through hierarchical integration rather than operating as independent
modules. The main contributions are summarized as follows:

\begin{itemize}

\item 
We propose a safety-critical multi-agent MCTS framework for 
dual-lane roundabouts, unifying safety pruning, uncertainty 
modeling, and reward shaping into an interaction-aware tree search 
pipeline for mixed traffic.

\item We develop a structurally differentiated safety assessment 
mechanism that handles various interactions via 
lane-dependent risk metrics and conservative thresholds for 
human uncertainty, validated through ablation studies.

\item We introduce two key structural components for robust 
coordination: (i) an adaptive HDV prediction model capturing 
lane-specific uncertainty and exit proximity effects, and 
(ii) a Shapley value-based reward allocation scheme enabling 
contribution-aware credit assignment and reducing collision rates 
in cooperative roundabout navigation.

\end{itemize}

\section{Safety-Critical Decision-Making Framework}\label{sec2}

We consider unsignalized dual-lane roundabouts, where CAVs must navigate among other CAVs and HDVs without traffic signals. This task requires handling both predictable autonomous behaviors and uncertainty in human driving patterns.

To model this interaction-rich environment, we cast the problem as a multi-agent Markov Decision Process (MDP), formally defined as $\langle \mathcal{S}, \mathcal{A}, \mathcal{T}, \mathcal{R}, \gamma \rangle$. Here, $\mathcal{S}$ represents the joint state space capturing critical vehicle-level information such as positions, velocities, and heading angles; $\mathcal{A}$ defines the joint action space, including bounded control inputs for acceleration and steering; $\mathcal{T}$ describes the transition dynamics; $\mathcal{R}$ is the reward function guiding agent behavior; and $\gamma$ is the discount factor used for long-term planning.

The main complexity arises from three intertwined interaction types: C2C, C2H, and C2B interactions. Within this framework, the goal is to determine an optimal policy $\pi^*$ that ensures both safety and efficiency in navigation. Denoting $\pi \in \Pi$ as a policy in the admissible space, $s_t$ as the system state at time step $t$, and $\mathcal{R}(s_t, \pi(s_t))$ as the instantaneous reward, the optimization problem can be expressed as:
\begin{equation}
\pi^* = \argmax_{\pi \in \Pi} \mathbb{E}\left[\sum_{t=0}^{T-1} \gamma^t \mathcal{R}(s_t, \pi(s_t))\right].
\end{equation}

An overview of our proposed framework is depicted in Fig.~\ref{fig2_toc}. The architecture is composed of four key modules. First, the multi-agent MDP model is established, including formal definitions of states, actions, and dynamic models with embedded safety constraints (Sec.~\ref{sec3a}). Next, we introduce a hierarchical safety mechanism that considers layered safety constraints across vehicle types and environmental boundaries, using dynamically updated safety thresholds and predictive risk assessment for HDVs (Sec.~\ref{sec3b}).

\begin{figure}[t]
    \centering
    \includegraphics[width=0.7\linewidth]{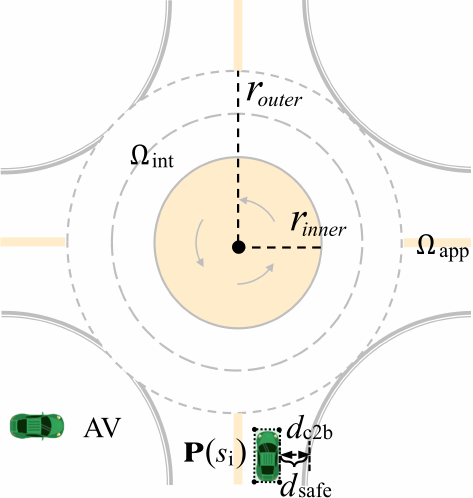}
    \vspace{-1mm}    
\caption{Illustration of the interaction scenario.}
    \label{figroad}
\end{figure}

Building on this foundation, we develop a safety-aware multi-agent 
MCTS algorithm with safety-encoded tree nodes, UCB-based 
exploration--exploitation balancing, and rollout-based policy 
extraction with backpropagation (Sec.~\ref{sec3c}). Finally, we 
formulate a multi-objective reward function that jointly accounts 
for A2A, A2H, and A2R safety, motion efficiency, and dynamic 
feasibility, yielding an integrated optimization scheme 
(Sec.~\ref{sec3d}).

Fig.~\ref{figroad} illustrates the interactive environment surrounding a CAV navigating a roundabout. The figure highlights key spatial regions that shape decision-making, including a central interaction zone where vehicles must account for potential conflicts and trajectory overlaps, and a broader approach area where speed and heading adjustments begin in anticipation of entering the roundabout. The inner circle represents the main circulation area, while the shaded zones near the exits denote heightened interaction regions where yielding, merging, and exiting are most critical. These regions collectively define the spatial context in which CAVs must coordinate with surrounding agents to ensure safe and efficient navigation, and the key notation is summarized in Table~\ref{tab:notation}.

\section{Safety-Critical Decision Making System}
\label{sec3}
\subsection{Multi-Agent MDP Formulation for Dual-Lane Roundabout}
\label{sec3a}
We formulate the dual-lane roundabout navigation problem involving $N$ CAVs and $M$ HDVs as a multi-agent MDP: $\langle \mathcal{S}, \mathcal{A}, \mathcal{T}, \mathcal{R}, \gamma \rangle$. For the $i$-th vehicle, the state vector $s_i$ consists of its polar coordinates $(r_i, \theta_i)$, velocity $v_i$, heading angle $\phi_i$, and lane index $l_i$. The control input $u_i$ includes acceleration command $a_i$, steering rate $\dot{\phi}_i$, and lane-changing decision $\delta_i$. The joint state and action spaces are defined as:
\begin{equation}
\mathcal{S} = \prod_{i=1}^{N+M} \left\{ s_i = 
\begin{bmatrix} 
r_i \\[2pt] 
\theta_i \\[2pt] 
v_i \\[2pt] 
\phi_i \\[2pt] 
l_i 
\end{bmatrix} \in \mathbb{R}^5
\;\middle|\;
\begin{aligned}
&r_i \in [r_{\text{inner}}, r_{\text{outer}}],\\[1pt]
&\theta_i \in [0, 2\pi),\\[1pt]
&v_i \in [0, v_{\max}],\\[1pt]
&\phi_i \in [-\pi, \pi],\\[1pt]
&l_i \in \{0, 1\}
\end{aligned}
\right\},
\end{equation}

\begin{table}[t]
\centering
\captionsetup{
    labelfont={sc}, 
    textfont={sc}, 
    labelsep=colon, 
    singlelinecheck=false,
    justification=centering,
    format=plain
}
\caption{Summary of Key Notation}
\label{tab:notation}
\setlength{\tabcolsep}{4pt}
\renewcommand{\arraystretch}{1.1}
\small
\begin{tabular}{cl}
\hline
\hline
Symbol & Description \\
\hline
\multicolumn{2}{l}{\footnotesize\textit{State and Control}} \\
$s_i = [r_i, \theta_i, v_i, \phi_i, l_i]^\top$ & State vector of vehicle $i$ \\
$u_i = [a_i, \dot{\phi}_i, \delta_i]^\top$ & Control input of vehicle $i$ \\
$\mathcal{C},\;\mathcal{H}$ & Set of CAVs / HDVs \\
$N,\;M$ & Number of CAVs / HDVs \\
\hline
\multicolumn{2}{l}{\footnotesize\textit{Safety Assessment}} \\
$d(s_i,s_j)$ & Min.\ distance between vehicles $i,j$ \\
$d_{\text{safe}}(s_i,s_j)$ & Adaptive safe distance \\
$d_{\min}$ & Min.\ base safe distance \\
$d_{\text{c2b}}(s_i)$ & Min.\ distance to roundabout boundary \\
$Q_{\text{risk}}^{\text{cc}},\;Q_{\text{risk}}^{\text{ch}}$ & C2C / C2H overall risk metric \\
$Q_{\text{th}}^{\text{cc}},\;Q_{\text{th}}^{\text{ch}}$ & Safety thresholds for C2C / C2H \\
$C_{ih}$ & C2H collision probability \\
\hline
\multicolumn{2}{l}{\footnotesize\textit{HDV Uncertainty Modeling}} \\
$f_{\text{IDM}}(s_h)$ & IDM nominal prediction for HDV $h$ \\
$\boldsymbol{\Sigma}_h^{\text{lane}}$ & Lane-specific covariance matrix \\
$\mathbf{M}_{\text{lane}}(l_h)$ & Lane multiplier matrix \\
$\mathbf{M}_{\text{exit}}(\mathbf{E}_h)$ & Exit proximity multiplier \\
\hline
\multicolumn{2}{l}{\footnotesize\textit{MCTS and Reward}} \\
$N_n,\;Q_n$ & Node visit count / cumulative reward \\
$e_{\text{exp}}$ & UCB exploration constant \\
$K$ & Number of MCTS iterations \\
$\phi_i$ & Shapley value of agent $i$ \\
$\lambda_i^t$ & Cooperation coefficient \\
\hline
\hline
\end{tabular}
\end{table}

\begin{equation}
\mathcal{A}_i = \left\{ u_i = 
\begin{bmatrix} 
a_i & \dot{\phi}_i & \delta_i 
\end{bmatrix}^{\top} \in \mathbb{R}^3
\;\middle|\;
\begin{aligned}
&|a_i| \leq a_{\max},\\[1pt]
&|\dot{\phi}_i| \leq \dot{\phi}_{\max},\\[1pt]
&\delta_i \in \{-1, 0, 1\}
\end{aligned}
\right\},
\end{equation}
where $r_{\text{inner}}$ and $r_{\text{outer}}$ are the inner and outer lane radii, $l_i = 0$ represents the inner lane, $l_i = 1$ represents the outer lane, and $\delta_i$ denotes the lane-changing decision (-1: move inward, 0: maintain, 1: move outward).
The roundabout navigation decisions must satisfy the following safety constraints on state transitions and inter-vehicle distances:
\begin{equation}
\label{eq5_roundabout}
\begin{aligned}
\mathcal{S}_{\text{safe}} &= \left\{\mathbf{s} \in \mathcal{S} \;\middle|\; d(s_i,s_j) \geq d_{\text{safe}}, \forall i,j \in \mathcal{C}\cup\mathcal{H} \right\}, \\[5pt]
d(s_i,s_j) &= \min_{\mathbf{p}_i \in \mathbf{P}(s_i),\, \mathbf{p}_j \in \mathbf{P}(s_j)} \|\mathbf{p}_i - \mathbf{p}_j\|_2,
\end{aligned}
\end{equation}
where $\mathbf{s} = [s_1, s_2, ..., s_{N+M}]^T$, $d(s_i,s_j)$ is the minimum distance between the $i$-th and $j$-th vehicles, $d_{\text{safe}}$ is the minimum safe distance, $\mathbf{P}(s_i)$ represents the four vertices of the $i$-th CAV as shown in Fig.~\ref{figroad}, $\mathcal{C}$ represents the set of CAVs and $\mathcal{H}$ represents the set of HDVs .
Let $\mathbf{s}_t = [s_{1,t}, s_{2,t}, ..., s_{N+M,t}]^T$ be the joint state vector of all vehicles at time step $t$, with $s_{i,t} = [r_{i,t}, \theta_{i,t}, v_{i,t}, \phi_{i,t}, l_{i,t}]^T$ being the state vector of vehicle $i \in \mathcal{C}\cup\mathcal{H}$.

Navigation decisions must also satisfy constraints over $T$:
\begin{equation}
\label{eq:state_evolve_roundabout}
\mathbf{s}{t+1} \in \mathcal{S}_{\text{safe}}, \forall t \in [0,T], |v_i| \leq v_{\max}, ~d_{\text{c2b}}(s_i) \geq d_{\min},
\end{equation}
where $d_{\text{c2b}}(s_i)$ is the minimum distance to roundabout boundaries, with the minimum allowable value $d_{\min}$.
The CAV state transitions can be precisely calculated using control inputs and dynamics as follows:
\begin{equation}
\label{dyna_cav}
\mathbf{s}_{t+1} = \boldsymbol{\Phi}(\mathbf{s}_t, \mathbf{u}_t),
\end{equation}
with the transition function $\boldsymbol{\Phi}(\mathbf{s}_t,\mathbf{u}_t) = {g_i(s_{i,t}, u_{i,t})}_{i=1}^N$. $\mathbf{u}_t = [u_{1,t}, u_{2,t}, \cdots, u_{N,t}]^T$ contains the control inputs for all CAVs at time $t$. $g_i(s_{i,t}, u_{i,t})$ contains the vehicle kinematic model of vehicle $i \in \mathcal{C}$ as follows:
\begin{equation}
\label{dyna_roundabout}
g_i(s_{i,t}, u_{i,t}) = 
\begin{bmatrix}
\text{sat}_{[r_{\text{inner}},\, r_{\text{outer}}]}\left(r_{i,t} + \frac{v_{i,t} \sin(\phi_{i,t})}{r_{i,t}} \Delta t\right) \\[5pt]
\text{wrap}_{[0,\, 2\pi]}\left(\theta_{i,t} + \frac{v_{i,t} \cos(\phi_{i,t})}{r_{i,t}} \Delta t\right) \\[5pt]
\text{sat}_{[0,\, v_{\max}]}\left(v_{i,t} + a_{i,t} \Delta t\right) \\[5pt]
\text{wrap}_{[-\pi,\, \pi]}\left(\phi_{i,t} + \dot{\phi}_{i,t} \Delta t\right) \\[5pt]
\text{LC}(l_{i,t},\, \delta_{i,t},\, \mathbf{Z}_{i,t})
\end{bmatrix},
\end{equation}
where $\Delta t$ is the time step, $\text{sat}{[a,b]}(\cdot)$ keeps values within bounds, $\text{wrap}{[a,b]}(\cdot)$ handles angle continuity, and $\text{LC}(l_{i,t}, \delta_{i,t}, \mathbf{Z}{i,t})$ is the lane-changing function that depends on current lane $l{i,t}$, decision $\delta_{i,t}$, and safety conditions $\mathbf{Z}_{i,t}$.
While CAV states are updated by \eqref{dyna_cav}, HDV behavior involves inherent uncertainties that grow over time and vary significantly between inner and outer lanes. These uncertainties must be properly characterized for reliable safety assessment in roundabout environments.

\begin{table}[t]
\centering
\setlength{\tabcolsep}{5pt}
\captionsetup{
    labelfont={sc}, 
    textfont={sc}, 
    labelsep=colon, 
    skip=1em,       
    singlelinecheck=false,
    justification=centering,
    format=plain
}
\caption{Safety distance adjustment factors}
\label{tab:safety_factors}
\begin{tabular}{cll}
\hline
Factor & Expression & Purpose \\
\hline
$\alpha_v$ & $1 + \beta_v|\Delta v_{ij}|/v_{\text{ref}}$ & Velocity adjustment \\
$\alpha_\phi$ & $1 + \beta_\phi|\Delta \phi_{ij}|/\pi$ & Heading adjustment \\
$\alpha_l$ & $1 + \beta_l|l_i - l_j|$ & Lane difference \\
$\alpha_z$ & $1 + \sum_{\Omega} \mathbb{1}_{\Omega}(s_i, s_j)$ & Zone-based risk \\
\hline
\end{tabular}
\end{table}

\subsection{Hierarchical Safety Assessment for Roundabout Navigation}
\label{sec3b}
To ensure safe navigation in dual-lane roundabouts, we develop a hierarchical safety assessment framework that evaluates three critical interaction types. For C2C interactions, we directly use vehicle states $s_i, s_j \in \mathcal{S}$ since their evolution can be precisely characterized by \eqref{dyna_roundabout}. For C2H interactions involving HDVs (denoted by $h \in \mathcal{H}$), we need to predict their future states $\hat{s}_h$ due to behavioral uncertainties. Additionally, C2B interactions are considered for spatial safety constraints within the roundabout geometry.

\textit{1) C2B safety assessment:}
The C2B safety assessment focuses on spatial constraints by partitioning the roundabout environment into the interaction area $\Omega_{\text{int}}$ and approach/exit areas $\Omega_{\text{app}}$:
\begin{equation}
\begin{aligned}
\Omega_{\text{int}} &= \left\{ s \in \mathcal{S} \;\middle|\; r_{\text{inner}} \leq r \leq r_{\text{outer}} \right\}, \\[5pt]
\Omega_{\text{app}} &= \left\{ s \in \mathcal{S} \;\middle|\; r < r_{\text{inner}} \;\vee\; r > r_{\text{outer}} \right\}.
\end{aligned}
\end{equation}

The safety level is evaluated through the minimum distance to roundabout boundaries $d_{\text{c2b}}(s_i)$ and its corresponding penalty function $\varphi_{\text{c2b}}(d)$, defined as:
\begin{equation}
\label{dc2b}
\begin{aligned}
d_{\text{c2b}}(s_i) &= \min_{p \in \mathbf{P}(s_i)} \text{distance}\left(p, \partial\Omega_{\text{circ}} \cup \partial\Omega_{\text{appr}}\right), \\[5pt]
\varphi_{\text{c2b}}(d) &= 
\begin{cases}
-\infty, & \text{if } d \leq d_{\min}, \\[3pt]
-\beta \left(\dfrac{d_{\min}}{d}\right)^2, & \text{if } d_{\min} < d \leq d_{\text{safe}}, \\[3pt]
0, & \text{if } d > d_{\text{safe}},
\end{cases}
\end{aligned}
\end{equation}
where $\partial$ denotes the area boundary and $\beta$ is a scaling factor.

\begin{figure}[t] 
    \centering
    \begin{subfigure}[c]{0.21\textwidth}
        \includegraphics[width=\linewidth]{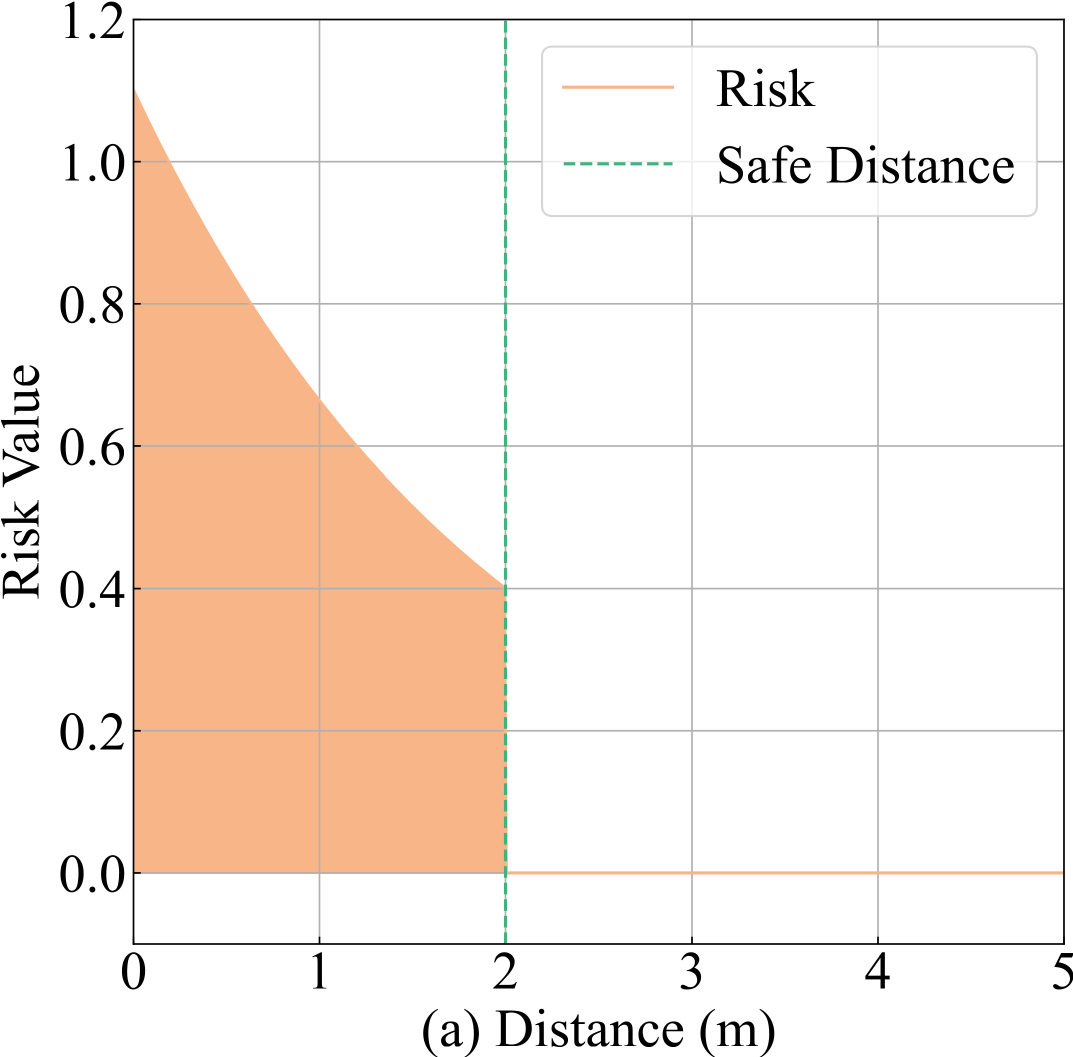}
        \label{fig11_3d_qulv_sub1}
    \end{subfigure}%
     \hspace{5pt} 
    \begin{subfigure}[c]{0.24\textwidth}
        \includegraphics[width=\linewidth]{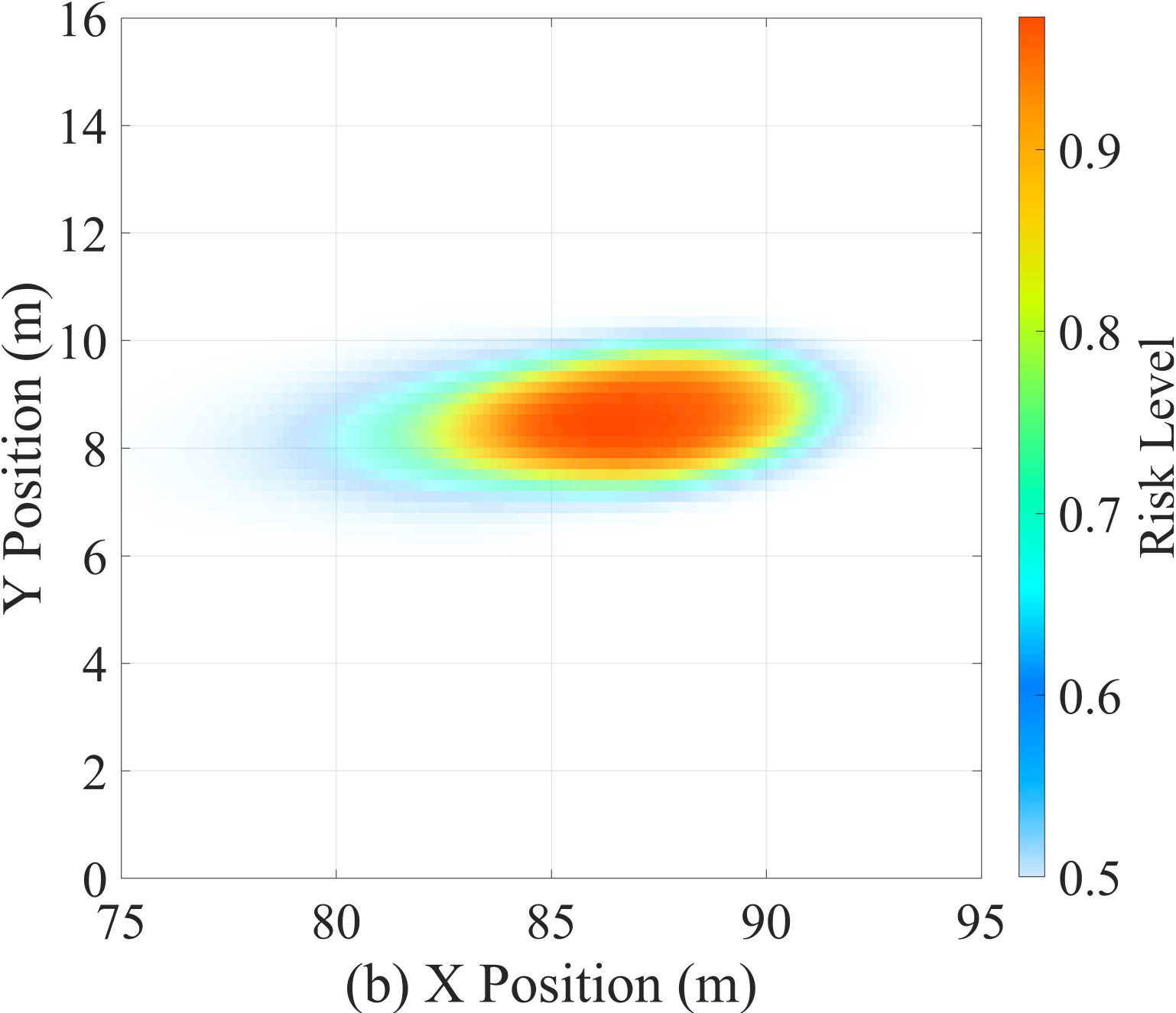}
        \label{fig11_3d_qulv_sub2}
    \end{subfigure}%
    \vspace{-3mm}
    \caption{Safety-critical risk assessment. (a) Distance-based risk. (b) safety risk visualization.}
    \label{fig3}
\end{figure}

\textit{2) C2C and C2H safety assessment:}
The adaptive safety distance accounts for multiple risk factors via a base value and adjustment terms, as shown in Table~\ref{tab:safety_factors}:

\begin{equation}
\label{eq:dsafe}
d_{\text{safe}}(s_i, s_j) = \max\{d_{\min}, \kappa_v|\Delta v_{ij}|\} \cdot \prod_{k \in \mathcal{K}} \alpha_k(s_i, s_j),
\end{equation}
where $d_{\min}$ is the minimum base distance, $\kappa_v$ is the velocity scaling factor, $\Delta v_{ij} = v_i - v_j$ is the relative velocity, $\Delta \phi_{ij}$ is the relative heading angle, $l_i, l_j \in \{0, 1\}$ denote lane indices, $v_{\text{ref}}$ is the reference velocity for normalization, $\beta_k$ are weighting parameters, and $\mathbb{1}_{\Omega}(\cdot)$ is the indicator function for vehicles in zone $\Omega$.  $\mathcal{K}=\{v,\phi,l,z\}$ indexes the adjustment factors in Table~\ref{tab:safety_factors}. Fig.~\ref{fig3} visualizes our safety-critical risk assessment framework for CAVs and HDVs navigating roundabouts. Fig.~\ref{fig3}(a) presents the foundational distance-based risk function, characterized by an exponential decay curve, which quantifies collision urgency and informs immediate avoidance actions in MCTS decisions. The clearly defined safe distance threshold enables effective and computationally efficient binary safety decisions. Fig.~\ref{fig3}(b) provides a spatial risk heat map for a representative roundabout scenario, highlighting high-risk zones at lane-changing and exit-merging points.

Based on the distance measure in \eqref{eq5_roundabout} and safety threshold in \eqref{eq:dsafe}, we define two complementary risk metrics. The instantaneous risk captures immediate collision threats:
\begin{equation}
\label{eq:risk_inst}
r_{\text{inst}}(s_i, s_j) = \exp\left(\frac{d_{\text{safe}} - d_{\min}}{  d_{\text{safe}}}\right) \cdot \left(1 +  \frac{|\Delta v_{ij}|}{v_{\max}}\right),
\end{equation}
where $d_{\min}$ is the minimum distance between vehicles and the exponential term penalizes proximity violations while the velocity term accounts for relative speed risks. The temporal risk aggregates future collision probabilities with time discounting:
\begin{equation}
\label{eq:risk_temp}
R_T(s_i, s_j) = \frac{1}{T}\sum_{t=1}^T \frac{1}{1+t} \cdot \rho\left(d_t, d_{\text{safe}}\right),
\end{equation}
where $\rho(d, d_{\text{safe}}) = \max\{0, (1 - d/d_{\text{safe}})^2\}$ provides a quadratic penalty when $d < d_{\text{safe}}$. For C2C interactions, the overall safety level is quantified by:
\begin{equation}
\label{qc2c}
Q_{\text{risk}}^{\text{cc}}(s_i, s_j) = w_1^{\text{cc}} r_{\text{inst}}(s_i, s_j) + w_2^{\text{cc}} R_T(s_i, s_j),~ i,j \in \mathcal{C},
\end{equation}
where $w_1^{\text{cc}}$ and $w_2^{\text{cc}}$ are given weights.

\textit{3) Lane-Specific HDV Uncertainty Modeling:}
For C2H interactions, we adopt the IDM~\cite{treiber2000congested} as the nominal behavior predictor and embed it within a probabilistic framework to account for behavioral deviations. The lane-specific covariance $\boldsymbol{\Sigma}_h^{\text{lane}}$ captures the distinct uncertainty profiles of inner and outer lanes, where drivers exhibit different levels of yielding tendency and speed variability. The uncertainty model is:
\begin{equation}
\label{hdv_predict_roundabout}
P(\hat{s}_h|s_h) = \mathcal{N}(f_{\text{IDM}}(s_h), \boldsymbol{\Sigma}_h^{\text{lane}}(t, l_h, \mathbf{E}_h)),
\end{equation}
where $f_{\text{IDM}}(s_h)$ captures the nominal human driving behavior,$s_h$ is the current state of HDVs, and $\boldsymbol{\Sigma}_h^{\text{lane}}(t, l_h, \mathbf{E}_h)$ is the lane-specific time-varying covariance matrix:
\begin{equation}
\boldsymbol{\Sigma}_h^{\text{lane}}(t, l_h, \mathbf{E}_h) = \boldsymbol{\Sigma}_{\text{base}}(t) \cdot \mathbf{M}_{\text{lane}}(l_h) \cdot \mathbf{M}_{\text{exit}}(\mathbf{E}_h),
\end{equation}
where the lane-specific multiplier matrix is:
\begin{equation}
\mathbf{M}_{\text{lane}}(l_h = 0) = \text{diag}(0.3, 0.5, 0.4, 0.6, 0.2),
\end{equation}
\begin{equation}
\mathbf{M}_{\text{lane}}(l_h = 1) = \text{diag}(1.2, 1.5, 1.3, 1.4, 1.8).
\end{equation}
The scaling factors are motivated by the geometric 
asymmetry of roundabout lanes: inner-lane vehicles are constrained 
by a tighter turning radius that naturally limits lateral and heading 
deviations, whereas outer-lane vehicles have greater freedom for 
radial displacement and exit maneuvers~\cite{11051098}. 
The effectiveness of this lane-specific design over uniform 
uncertainty is validated in the ablation study 
(Section~\ref{sec:sensitivity_ablation}).

The exit proximity multiplier $\mathbf{M}_{\text{exit}}(\mathbf{E}_h)$ accounts for increased uncertainty near exits:
\begin{equation}
\begin{aligned}
\mathbf{M}_{\text{exit}}(\mathbf{E}_h) &= \mathbf{I} + \sum_{k=1}^{N{\text{exit}}} \xi_k \exp\left(-\frac{|\theta_h - \theta_{\text{exit},k}|^2}{2\sigma_{\text{exit}}^2}\right) \mathbf{J}_k,
\end{aligned}
\end{equation}
where $N_{\text{exit}}$ is the number of exits, $\theta_{\text{exit},k}$ is the angular position of the $k$-th exit, $\xi_k$ is the exit influence factor, $\sigma_{\text{exit}}$ is the exit influence range, and $\mathbf{J}_k$ is a diagonal matrix containing state-specific uncertainty amplification factors for the $k$-th exit.
The base covariance matrix $\boldsymbol{\Sigma}_{\text{base}}(t)$ is defined as:
\begin{equation}
\begin{aligned}
&\boldsymbol{\Sigma}_{\text{base}}(t) = \\
&\begin{bmatrix}
\sigma_r^2 t + \epsilon_r^2 t^2 &\!\! 0 &\!\! \rho_{rv} \sigma_r \sigma_v t &\!\! 0 &\!\! 0 \\
0 &\!\! \sigma_\theta^2 t + \epsilon_\theta^2 t^2 &\!\! 0 &\!\! \rho_{\theta\phi} \sigma_\theta \sigma_\phi t &\!\! 0 \\
\rho_{rv} \sigma_r \sigma_v t &\!\! 0 &\!\! \sigma_v^2 &\!\! 0 &\!\! 0 \\
0 &\!\! \rho_{\theta\phi} \sigma_\theta \sigma_\phi t &\!\! 0 &\!\! \sigma_\phi^2 &\!\! 0 \\
0 &\!\! 0 &\!\! 0 &\!\! 0 &\!\! \sigma_l^2
\end{bmatrix},
\end{aligned}
\end{equation}
where $(\sigma_{r,\theta}, \epsilon_{r,\theta})$ capture position uncertainty growth including the radial uncertainty measured by $\sigma_r^2 t + \epsilon_r^2$ and angular uncertainty measured by $\sigma_\theta^2 t + \epsilon_\theta^2 t^2$ , while $(\sigma_{v,\phi}, \rho_{rv}, \rho_{\theta\phi})$ model velocity, heading, and state correlations. 

\begin{figure}[t]
    \centering
    \includegraphics[width=1\columnwidth]{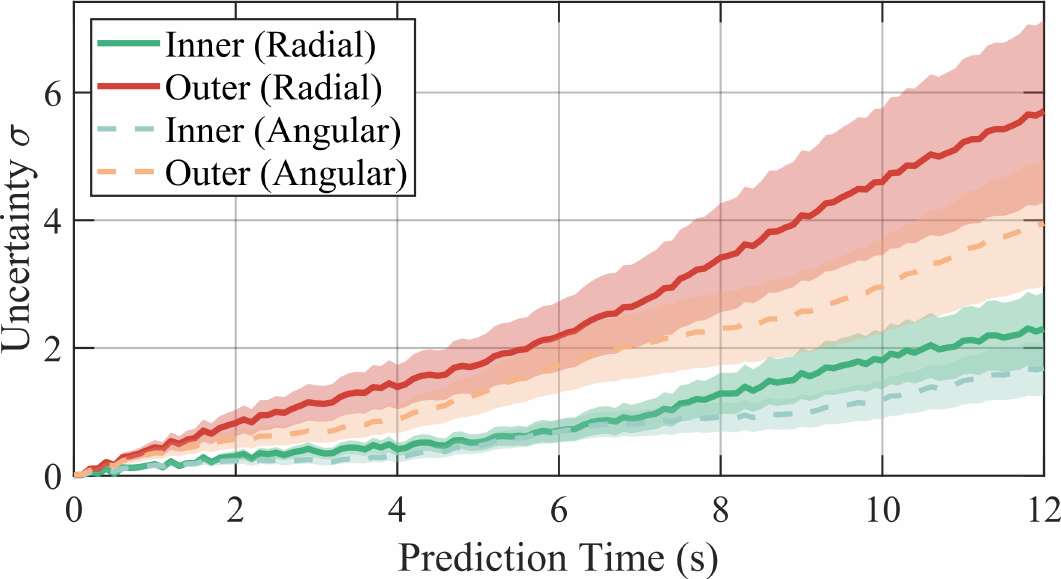}
    \caption{Oscillating temporal uncertainty evolution showing the growth of prediction uncertainty over time for inner and outer lanes in both radial and angular dimensions.}
    \label{fig:hdv_uncertainty}
\end{figure}

Fig.~\ref{fig:hdv_uncertainty} illustrates the temporal evolution of prediction uncertainty for HDVs in inner and outer lanes. Over a 12-second horizon, outer-lane radial uncertainty (red line) reaches approximately 5.8$\sigma$ compared to 2.5$\sigma$ for the inner lane (green line), while angular uncertainties show similar disparities (4.0$\sigma$ outer vs. 2.0$\sigma$ inner). The shaded regions represent 75\%--125\% confidence envelopes. This lane-specific uncertainty growth validates our hierarchical safety framework, where outer-lane vehicles require larger safety margins due to their less constrained movement patterns.

We bound HDVs' reachable state space as:
\begin{equation}
\begin{aligned}
\mathcal{S}_h^t = \left\{ \hat{s}_h \in \mathbb{R}^5 ~\middle|~ 
\begin{aligned}
&\|\mathbf{p}_h - \mathbf{p}_h(t)\| \leq (v_{\max} + \sigma_v)t, \\
&|v_h| \leq v_{\max} + 2\sigma_v, \\
&|\phi_h| \leq \pi, \\
&l_h \in \{0, 1\}
\end{aligned}
\right\}.
\end{aligned}
\end{equation}
The collision probability for C2H interactions is:
\begin{equation}
\label{colli_prob_roundabout}
\begin{aligned}
C_{ih} = \int_{\hat{s}_h \in \mathcal{S}_h^t} \psi(\hat{s}_i,\hat{s}_h) \cdot \mathcal{N}(f_{\text{IDM}}(s_h), \boldsymbol{\Sigma}_h^{\text{lane}})\mathrm{d}\hat{s}_h,
\end{aligned}
\end{equation}
where $\psi(\hat{s}_i,\hat{s}_h) = \mathbb{I}(d(\hat{s}_i,\hat{s}_h) < d{\text{safe}}(s_i, s_h))$ is the collision indicator function.
The safety level for C2H interactions is:
\begin{equation}
\label{qc2h}
Q_{\text{risk}}^{\text{ch}}(s_i, s_h) \!= w_1^{\text{ch}}  r_{\text{inst}}(s_i, \hat{s}_h) 
\!+\! w_2^{\text{ch}} R_T(s_i, s_h) 
\!+\! w_3^{\text{ch}} C_{ih},
\end{equation}
where$ i \in \mathcal{C},~ h \in \mathcal{H}$, and $w_k^{\text{ch}}, k \in [1,3]$ are given weights.

\section{Multi-Agent MCTS Solution Approach}
\label{sec35}
\subsection{Multi-Agent MCTS for Roundabout Navigation}
\label{sec3c}
Building on the safety assessment framework, we propose a structured tree search approach where the risk assessment functions in \eqref{qc2c} and \eqref{qc2h} are used to evaluate safety at each node and prune unsafe nodes that exceed predefined safety thresholds ($Q_{\text{th}}^{\text{cc}}$ for C2C interactions and $Q_{\text{th}}^{\text{ch}}$ for C2H interactions).
Let $\mathcal{T}$ be the search tree whose node $n \in \mathcal{T}$ is defined as:
$
n = (d_n,\, p_n,\, C_n,\, N_n,\, Q_n,\, u_n,\, \xi_n). 
$
where $d_n$ is the node depth, $p_n$ is the parent node, $C_n$ is the set of child nodes, $N_n$ is the visit count, $Q_n$ is the cumulative reward value, and $u_n$ is the action leading to this node.
A node is considered safe ($\xi_n = \text{safe''}$) if the safety metrics satisfy: $Q_{\text{risk}}^{\text{cc}} \leq Q_{\text{th}}^{\text{cc}}$ for C2C interactions, $Q_{\text{risk}}^{\text{ch}} \leq Q_{\text{th}}^{\text{ch}}$ for C2H interactions, and $d_{\text{c2b}} \geq d_{\min}$. Otherwise, it is marked as unsafe ($\xi_n = \text{unsafe''}$).
The joint policy space $\mathcal{A}_{\text{joint}} = \bigotimes{i=1}^{N+M} \mathcal{P}_i$ represents the Cartesian product of policy sets.

We set the safety thresholds to 
$Q_{\text{th}}^{\text{cc}} = 0.8$ (C2C) and 
$Q_{\text{th}}^{\text{ch}} = 0.6$ (C2H), 
using a more conservative margin for human-driven interactions 
due to higher prediction uncertainty. 
These values are calibrated in preliminary experiments to balance 
safety and traffic efficiency, and the sensitivity to $d_{\min}$ is 
further studied in Section~\ref{sec:sensitivity_ablation}. 
Following standard surrogate safety analysis, we adopt a PET 
threshold of $1.0$\,s~\cite{11227114}.

The safety-critical multi-agent MCTS algorithm uses a depth-first search strategy with safety validation at each node expansion. The search process evaluates nodes recursively:
\begin{equation}
\text{Search}(n_t) = 
\begin{cases}
\text{Expand}(n_t) \cup \text{Rollout}(n_t), & \text{if new \& safe}, \\[4pt]
\text{Search}(\text{UCB}(n_t)), & \text{if visited}, \\[4pt]
\text{Terminate}, & \text{if unsafe}.
\end{cases}\notag
\end{equation}

\begin{figure}[t]
    \centering
    \includegraphics[width=0.9\linewidth]{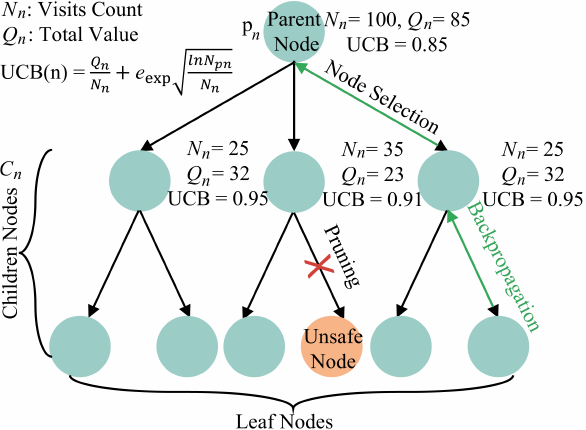}
    \vspace{-1mm}  
\caption{Illustration of the safety-critical MCTS framework. Each node stores the visit count $N_n$, value estimate $Q_n$, and UCB score. Nodes identified as unsafe (shown in orange) are pruned during the safety validation stage. The green path indicates the backpropagation process.}
\label{mstc}
\end{figure}

The selection of nodes for expansion is governed by the UCB formula, which balances exploration and exploitation:
\begin{equation}
\text{UCB}(n) = \frac{Q_n}{N_n} + e_{\text{exp}} \sqrt{\frac{\ln N_{p_n}}{N_n}},    
\end{equation}
where $e_{\text{exp}}$ is the exploration constant balancing exploitation and exploration, and $N_{p_n}$ is the visit count of the parent node. Fig.~\ref{mstc} illustrates the structure of an MCTS guided by the UCB strategy. Leaf nodes represent terminal or intermediate states, with one marked as an ``Unsafe Node'' indicating trajectories that violate safety constraints.

\subsection{Roundabout-Specific Reward Function Design}
\label{sec3d}

Roundabout navigation involves dense, asymmetric interactions that require both individual planning and group-level coordination. To accommodate these challenges, we propose a structured reward function that integrates four key aspects: safety, efficiency, comfort, and lane-specific behavior.

\textit{1) Group-aware reward formulation} We adopt a coalition-based design using the Shapley value $\phi_i$ to distribute the global reward among cooperative vehicles:
\begin{equation}
\phi_i(v) = \sum_{c\subset\mathcal{C}\cup\mathcal{H},i\in c} \frac{(|c| - 1)!(n - |c|)!}{n!}[v(c) - v(c \setminus \{i\})],\notag
\end{equation}
where $v(c) = \sum_{j \in c} \mathcal{R}_j(\mathbf{s}_t, \mathbf{u}_t)$ denotes the coalition value at time $t$ and $v(c \setminus \{i\})$ denotes the coalition value excluding CAV$_i$, such that the difference $v(c) - v(c \setminus \{i\})$ reflects how much value agent $i$ brings to coalition $c$. The coefficient $\frac{(|c| - 1)!(n - |c|)!}{n!}$ represents the probability that CAV$_i$ is the last to join group $c$ when CAVs join in a random order. This helps average $i$'s contribution across all possible team formations.

The global reward at each timestep is then defined as:
\begin{equation}
\mathcal{R}(\mathbf{s}_t, \mathbf{u}_t) = \sum_{i \in \mathcal{C}} \phi_i \cdot \mathcal{R}_i(\mathbf{s}_t, \mathbf{u}_t).
\end{equation}

Each CAV’s individual reward $\mathcal{R}_i$ balances its own performance and the impact on others:
\begin{equation}
\mathcal{R}_i = \frac{Q_i^{\text{self}} + \lambda_i^t Q_i^{\text{other}}}{1 + \lambda_i^t (N - 1)},
\end{equation}
where $Q_i^{\text{self}} = Q_{\text{safety}}^i + Q_{\text{eff}}^i + Q_{\text{comfort}}^i+Q_{\text{lane}}^i $, and $Q_i^{\text{other}} = \sum_{j \neq i} (Q_{\text{safety}}^j + Q_{\text{eff}}^j+Q_{\text{lane}}^j )$. The cooperation coefficient $\lambda_i^t \in [0,1]$ is fixed for CAVs and  empirical estimates  based on typical driving patterns for HDVs (e.g., $\lambda_i^t = 0.4$).

\textit{2) Reward components:} Each CAV's self-reward comprises four components that balance safety, efficiency, comfort, and lane-specific behaviors:
\begin{equation}
\begin{aligned}
Q_{\text{safety}}^i &= -w_{\text{c2c}} Q_{\text{risk}}^{\text{cc}}(s_i) - w_{\text{c2h}} Q_{\text{risk}}^{\text{ch}}(s_i) - w_{\text{c2b}} \varphi_{\text{c2b}},\\
Q_{\text{eff}}^i &= -w_v(v_i - v_{\text{des}})^2 - w_p\|p_i - p_{\text{ref}}\|^2,\\
Q_{\text{comfort}}^i &= -w_a|a_i|^2 - w_\phi|\dot{\phi}_i|^2,\\
Q_{\text{lane}}^i &= Q_{\text{position}}^i(l_i) - w_{\text{trans}}|\delta_i|^2 + Q_{\text{exit}}^i(\theta_i),
\end{aligned}
\end{equation}
where the safety terms directly incorporate the risk metrics from \eqref{dc2b}, \eqref{qc2c} and \eqref{qc2h}, with $\varphi_{\text{c2b}}$ penalizing boundary violations. The efficiency component encourages desired velocity $v_{\text{des}}$ and path tracking to reference trajectory $p_{\text{ref}}$. The comfort term penalizes excessive acceleration $a_i$ and yaw rate $\dot{\phi}_i$.

\textit{3) Summary} This hierarchical reward design enables agents to balance local objectives with socially-aware behavior in the roundabout, while explicitly incorporating safety, comfort, and intent-awareness. The cooperation coefficient $\lambda_i$ modulates each agent’s autonomy versus altruism, and the Shapley-based allocation ensures fair credit assignment in joint outcomes.

\subsection{Optimization Policy}
For HDVs in the roundabout, we use the IDM model with lane-specific parameters as described in Section \ref{sec3b}. During MCTS rollouts, HDV actions are sampled from the predictive distribution in \eqref{hdv_predict_roundabout}, with cooperation levels $\lambda_h$ based on typical driving patterns. 
The optimality of our MCTS solution for roundabout navigation can be formally characterized by the following optimization objective:
\begin{equation}
\mathbf{u}^* = \argmax_{\mathbf{u} \in \mathcal{A}_{\text{joint}}} \mathbb{E}\left[\sum_{t=0}^{T-1} \gamma_{r}^t \sum_{i \in \mathcal{C}} \phi_i\mathcal{R}_i(\mathbf{s}_t, \mathbf{u}(\mathbf{s}_t)) \middle| \mathbf{u}\right],
\end{equation}
subject to the following constraints at each time step $t$:
\begin{equation}
\begin{array}{ll}
\textbf{Safety Constraints:} & \\
\quad \text{(C2C)}\quad 
Q_{\text{risk}}^{\text{cc}}(s_i, s_j) \leq Q_{\text{th}}^{\text{cc}}, 
& \forall i,j \in \mathcal{C} \\[4pt]
\quad \text{(C2H)}\quad 
Q_{\text{risk}}^{\text{ch}}(s_i, s_h) \leq Q_{\text{th}}^{\text{ch}}, 
& \forall i \in \mathcal{C},\, h \in \mathcal{H} \\[4pt]
\quad \text{(C2B)}\quad 
d_{\text{c2b}}(s_i) \geq d_{\min}, 
& \forall i \in \mathcal{C} \\[8pt]

\textbf{Dynamic Constraints:} & \\
\quad \text{(Velocity)}\quad 
v_i \in [0, v_{\max}], 
& \forall i \in \mathcal{C} \\[4pt]
\quad \text{(Acceleration)}\quad 
|a_i| \leq a_{\max}, 
& \forall i \in \mathcal{C} \\[4pt]
\quad \text{(Steering)}\quad 
|\dot{\phi}_i| \leq \dot{\phi}_{\max}, 
& \forall i \in \mathcal{C} \\[8pt]
\end{array}
\end{equation}
This formulation provides a comprehensive theoretical characterization of our roundabout navigation approach. The reward function $\mathcal{R}_i(\mathbf{s}t, \mathbf{u}(\mathbf{s}t))$  incorporates Shapley value-based contribution quantification. The lane-specific modeling through matrices $\mathbf{M}{\text{lane}}(l_h)$ and $\mathbf{M}{\text{exit}}(\mathbf{E}_h)$ enables precise characterization of uncertainty variations between inner and outer lanes, while the exit proximity effects provide realistic behavioral modeling for roundabout scenarios.
Following the kinematic model in \eqref{dyna_roundabout}, our framework addresses roundabout conflicts by integrating safety assessment, reward design, and MCTS-based policy search, ensuring safe and efficient navigation through the balance of individual and cooperative behaviors.


\subsection{Computational Complexity Analysis}

The computational complexity of the proposed multi-agent MCTS-based planning algorithm arises from three main components: joint-action tree expansion, safety validation, and rollout simulation. Tree expansion considers joint actions over $N$ CAVs, each with $|\mathcal{A}|$ possible actions, resulting in a branching factor of $|\mathcal{A}|^N$. At each expanded node, safety validation involves pairwise CAV interactions ($O(N^2)$), CAV-to-HDV checks ($O(NM)$), and lane-level uncertainty evaluations, typically contributing $O(NM \cdot d^2)$ complexity, where $d$ is the state dimension.

Each new node further triggers a rollout simulation up to a depth $d_{\max}$, with per-step computations dominated by state transitions, reward evaluations, and uncertainty updates, also scaling with $O(NM \cdot d^2)$.

As a result, the worst-case overall complexity across $K$ MCTS iterations can be coarsely estimated as:
\begin{equation}
O\left(K \cdot |\mathcal{A}|^N \cdot d_{\max} \cdot NM \cdot d^2\right),
\end{equation}
which grows exponentially with the number of agents.

In practice, however, the actual computational cost is significantly lower due to three factors: 1) safety pruning that eliminates unsafe branches early; 2) UCB-guided selective search that limits unnecessary expansions; and 3) matrix-based parallel implementation of safety and reward computations. While the lane-specific uncertainty modeling adds overhead, it substantially improves prediction robustness in congested or uncertain roundabout scenarios.

\begin{table}[t]
\centering
\captionsetup{
    labelfont={sc}, 
    textfont={sc}, 
    labelsep=colon, 
    singlelinecheck=false,
    justification=centering,
    format=plain
}
\setlength{\tabcolsep}{4pt}
\caption{Computational complexity comparison}
\label{tab:complexity}
\begin{tabular}{l|c|c}
\hline
\hline
Method & Complexity & Operations \\
\hline
Joint exhaustive search & $|\mathcal{A}|^{N \cdot H}$ 
  & $\approx 10^{85}$ \\
Game-theoretic Nash & $\mathcal{O}(|\mathcal{A}|^{NH} 
  \cdot I_{\text{Nash}})$ & $> 10^{87}$ \\
MCTS (no pruning) & $K \cdot |\mathcal{A}| \cdot N 
  \cdot H$ & $\approx 3.2 \times 10^{5}$ \\
\rowcolor{lightcyan}
\textbf{MCTS + safety pruning} & $(1-\rho) \cdot K 
  \cdot |\mathcal{A}| \cdot N \cdot H$ 
  & $\approx \mathbf{2.2 \times 10^{5}}$ \\
\hline
\hline
\end{tabular}
\end{table}

To provide practical runtime context, we report 
measured per-step computation times from our MATLAB implementation. 
In our most complex scenario (8-vehicle symmetric intersection), 
the framework plans for $N = 4$ decision-making CAVs while the 
remaining HDVs are modeled as reactive agents with negligible 
planning overhead. The average planning time is approximately 
58\,ms per decision step, well within a typical planning cycle of 
$\Delta t = 0.2$\,s. Given the linear scaling with $N$ in our 
MCTS framework, we estimate the current implementation can support 
approximately $N = 6$--$8$ planning agents in real time. As shown 
in Table~\ref{tab:complexity}, safety pruning reduces the effective 
search space by approximately $1.4\times$ compared to unpruned MCTS, 
while UCB-guided selection further concentrates search effort on 
promising branches. A C++ implementation with parallelized rollouts 
would further extend this scalability envelope, left for future work.

\begin{figure}[t]
    \centering
    \begin{subfigure}{\columnwidth}
    \centering
        \includegraphics[ width=\textwidth]{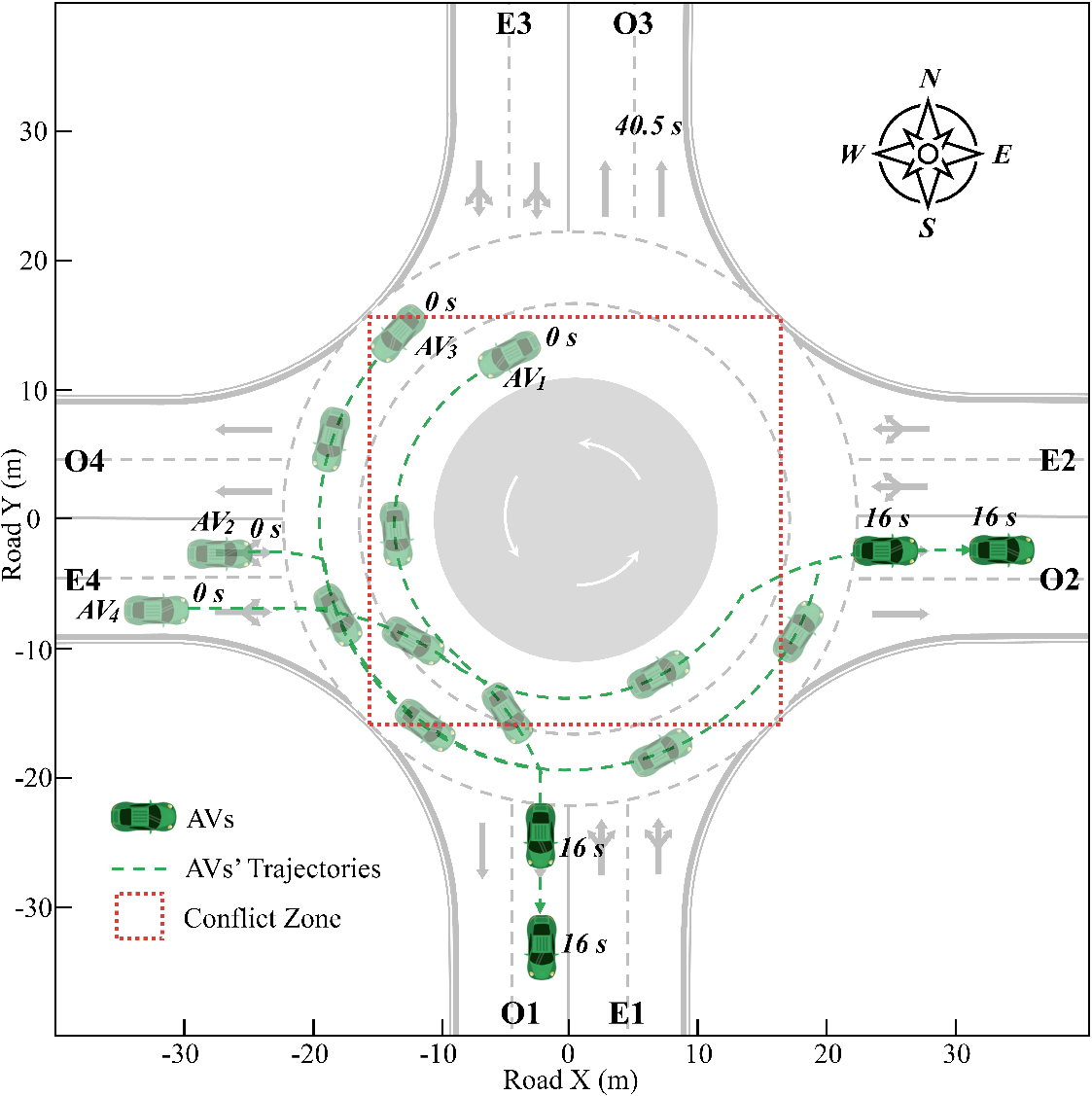}
         \vspace{-6mm}
        \caption{Illustration of MCTS simulation at a dual-lane roundabout.}
        \label{fig6_1}
    \end{subfigure}
 
\begin{subfigure}{\columnwidth}
\centering
    \includegraphics[width=0.95\textwidth]{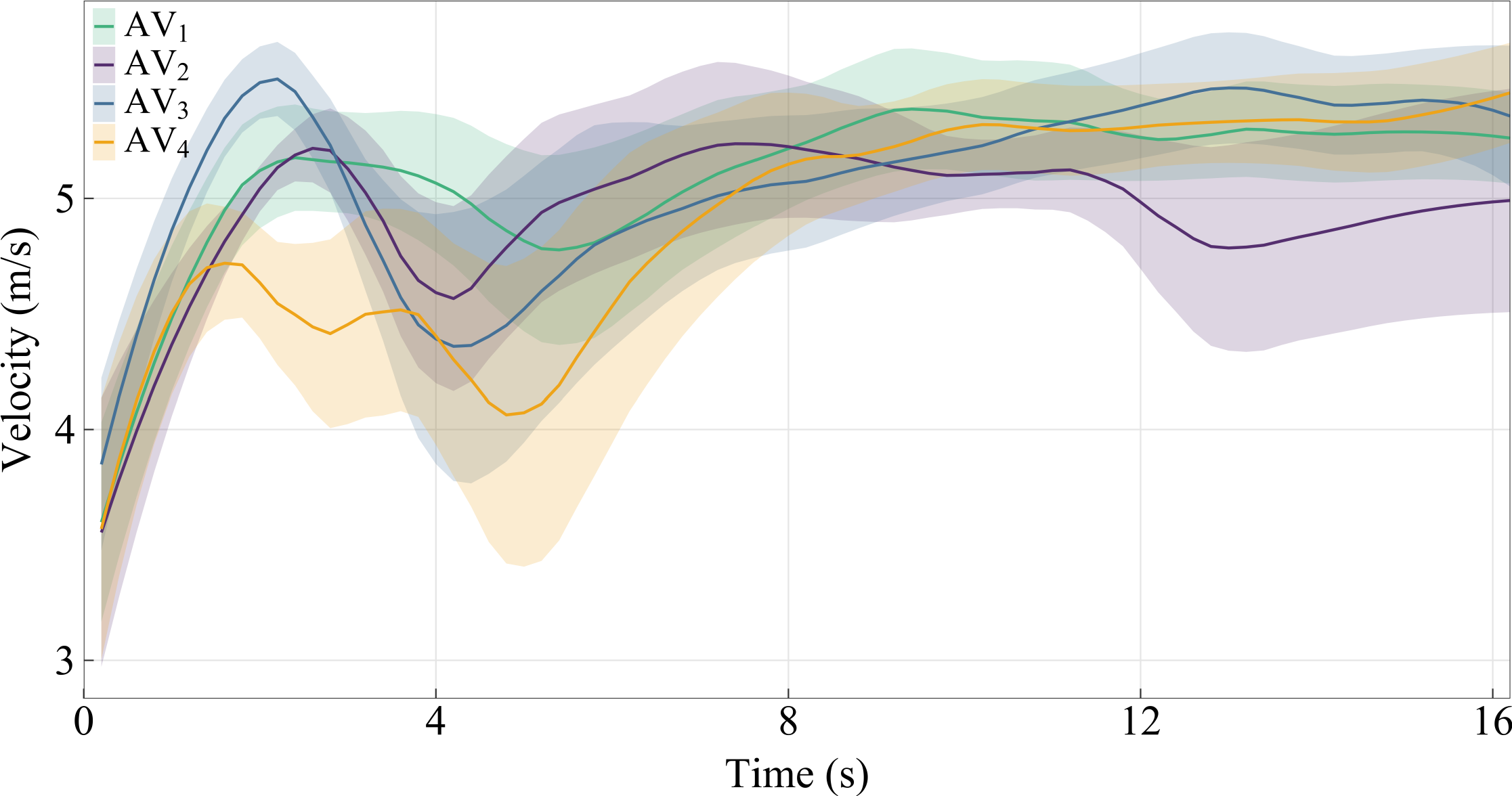}
    \vspace{-2mm}
    \caption{Velocity profiles of CAVs.}
    \label{fig6_2}
\end{subfigure}%

\begin{subfigure}{\columnwidth}
\vspace{-0mm}
\centering
    \includegraphics[width=0.95\textwidth]{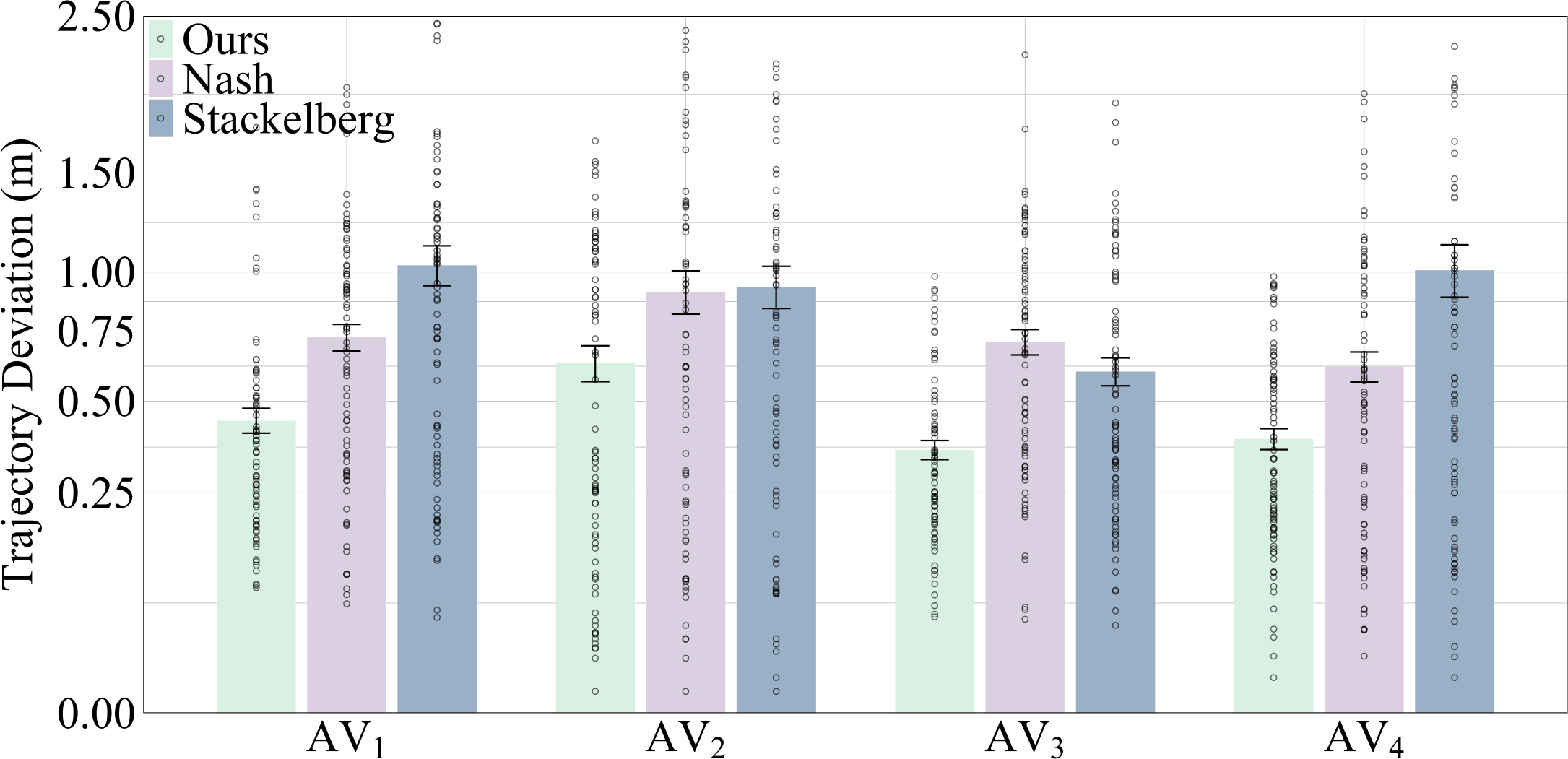}
    \vspace{-2mm}
    \caption{Comparison of trajectory deviations with advanced approaches.}
    \label{fig6_3}
\end{subfigure}%
     \vspace{-2mm}
    \caption{Performance Analysis in Case 1 (ROP = 100\%). }
    \label{fig6}
\end{figure}
\section{EXPERIMENTAL EVALUATION}
\label{sec4}
Simulations are conducted in MATLAB 2024a to evaluate the proposed approach for safe and efficient autonomous driving at a signal-free, dual-lane roundabout. We compare the proposed method with several advanced optimization algorithms, including the Stackelberg game approach \cite{hang2021decision} and the Nash equilibrium method \cite{9781400}.

\begin{figure}[t]
    \centering
    \begin{subfigure}{\columnwidth}
    \centering
        \includegraphics[ width=\textwidth]{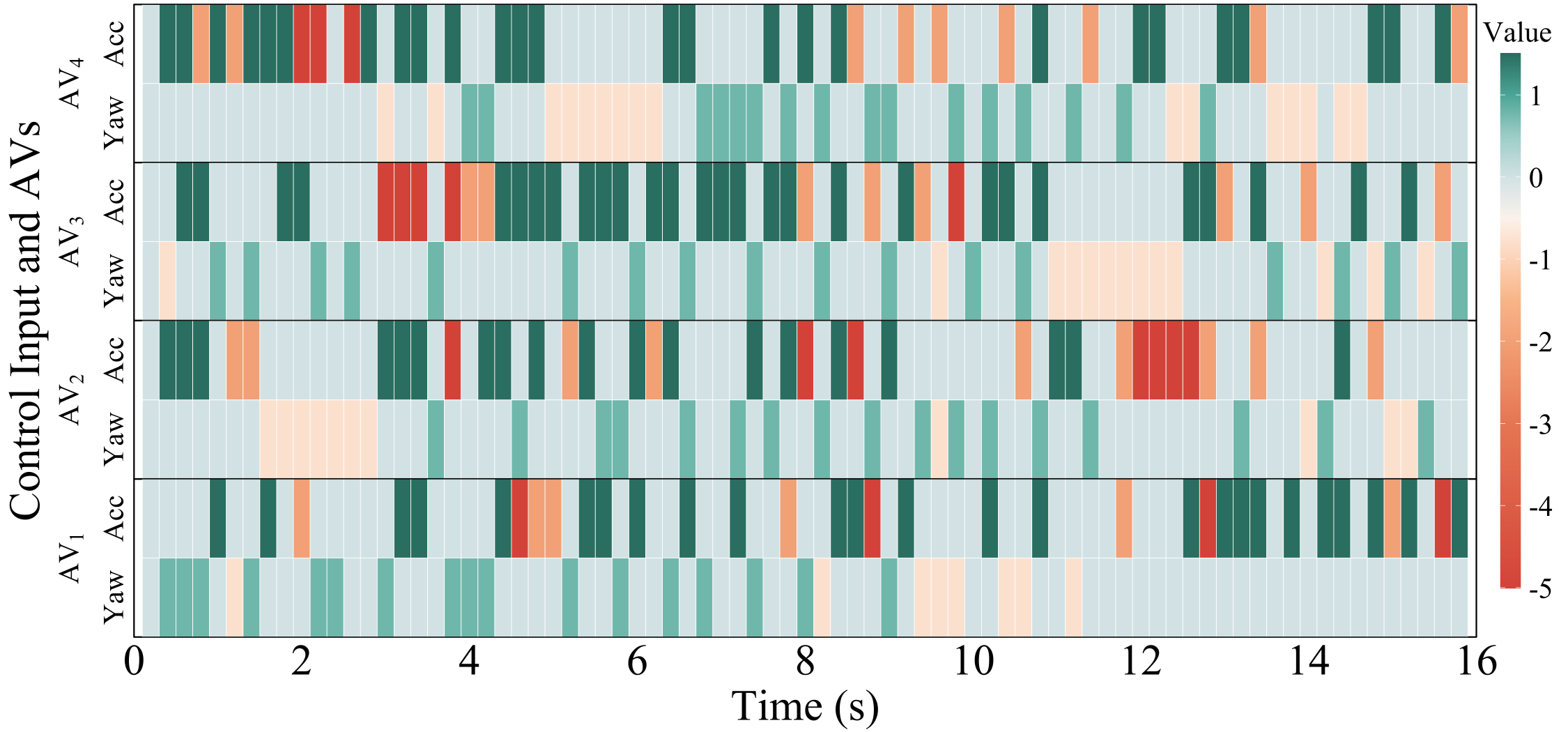}
         \vspace{-6mm}
        \caption{Heatmap of control inputs for controlled CAVs.}
        \label{fig7_1}
    \end{subfigure}
 
\begin{subfigure}{\columnwidth}
\centering
    \includegraphics[width=1\textwidth]{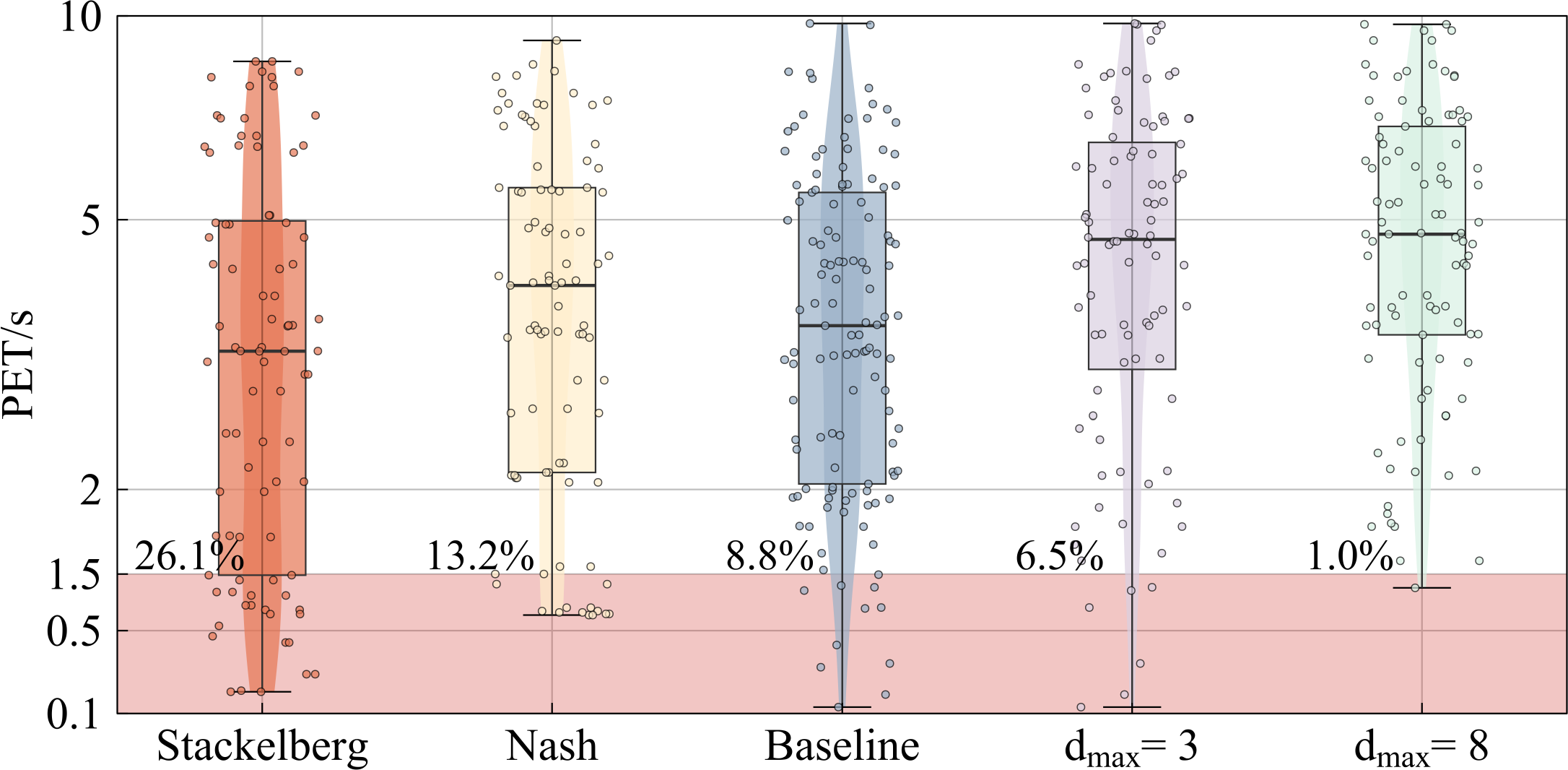}
    \vspace{-5mm}
    \caption{PET comparison.}
    \label{fig7_2}
\end{subfigure}%
     \vspace{-2mm}
    \caption{Analysis of decision-making and safety performance. (a) Variations in control inputs. (b) PET distributions and violations: benchmarks, Baseline (without adaptive risk evaluation), and our methods with $d_{\text{max}}$ of 3 and 8 (ROP = 100\%). }
    \label{fig7}
\end{figure}

\begin{figure}[t]
    \centering
    \begin{subfigure}{\columnwidth}
    \centering
        \includegraphics[ width=\textwidth]{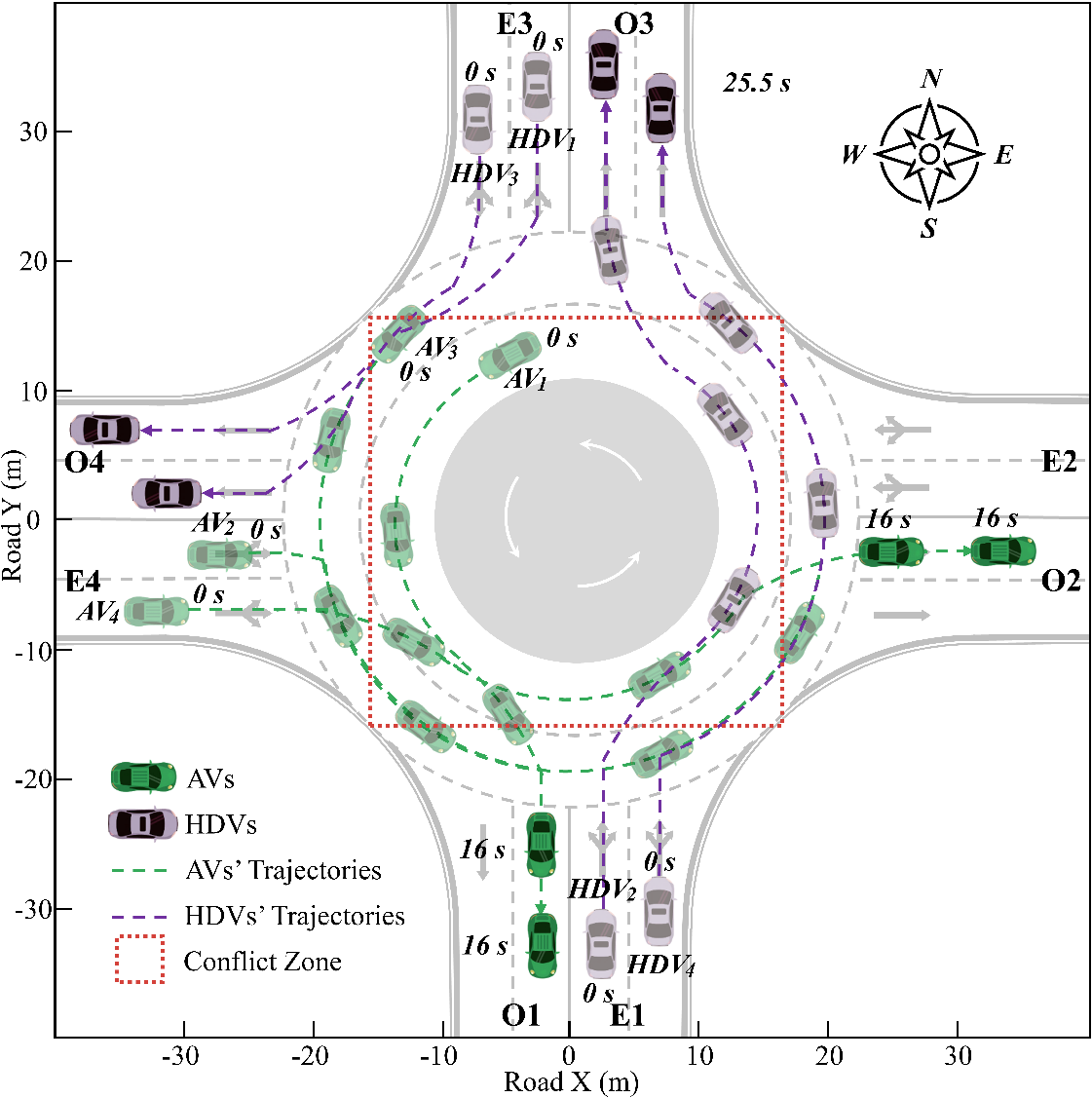}
         \vspace{-6mm}
        \caption{Illustration of MCTS simulation at a dual-lane roundabout.}
        \label{fig8_1}
    \end{subfigure}
 
\begin{subfigure}{\columnwidth}
\centering
    \includegraphics[width=0.95\textwidth]{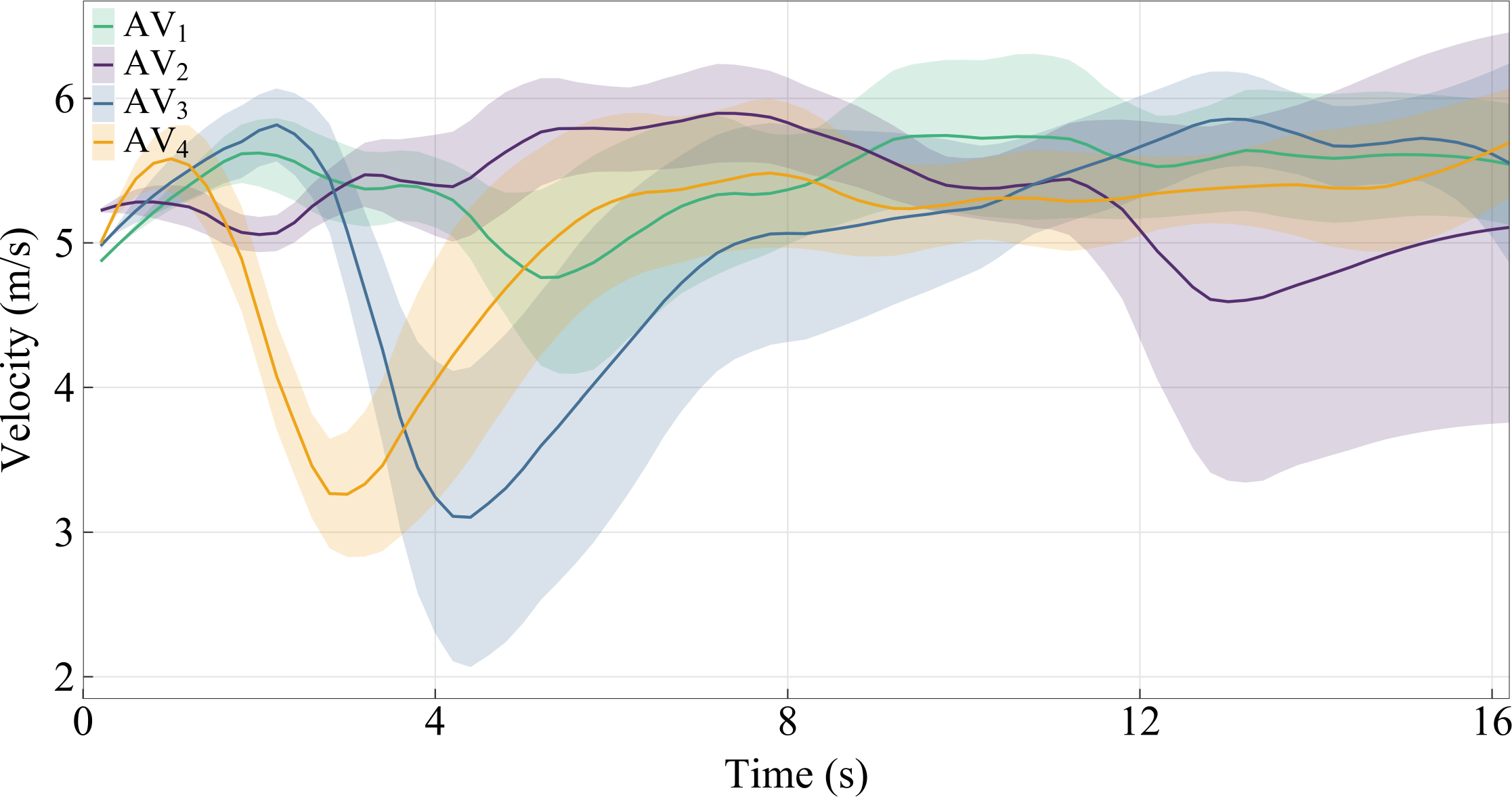}
    \vspace{-2mm}
    \caption{Velocity profiles of CAVs.}
    \label{fig8_2}
\end{subfigure}%

\begin{subfigure}{\columnwidth}
\vspace{-0mm}
\centering
    \includegraphics[width=0.95\textwidth]{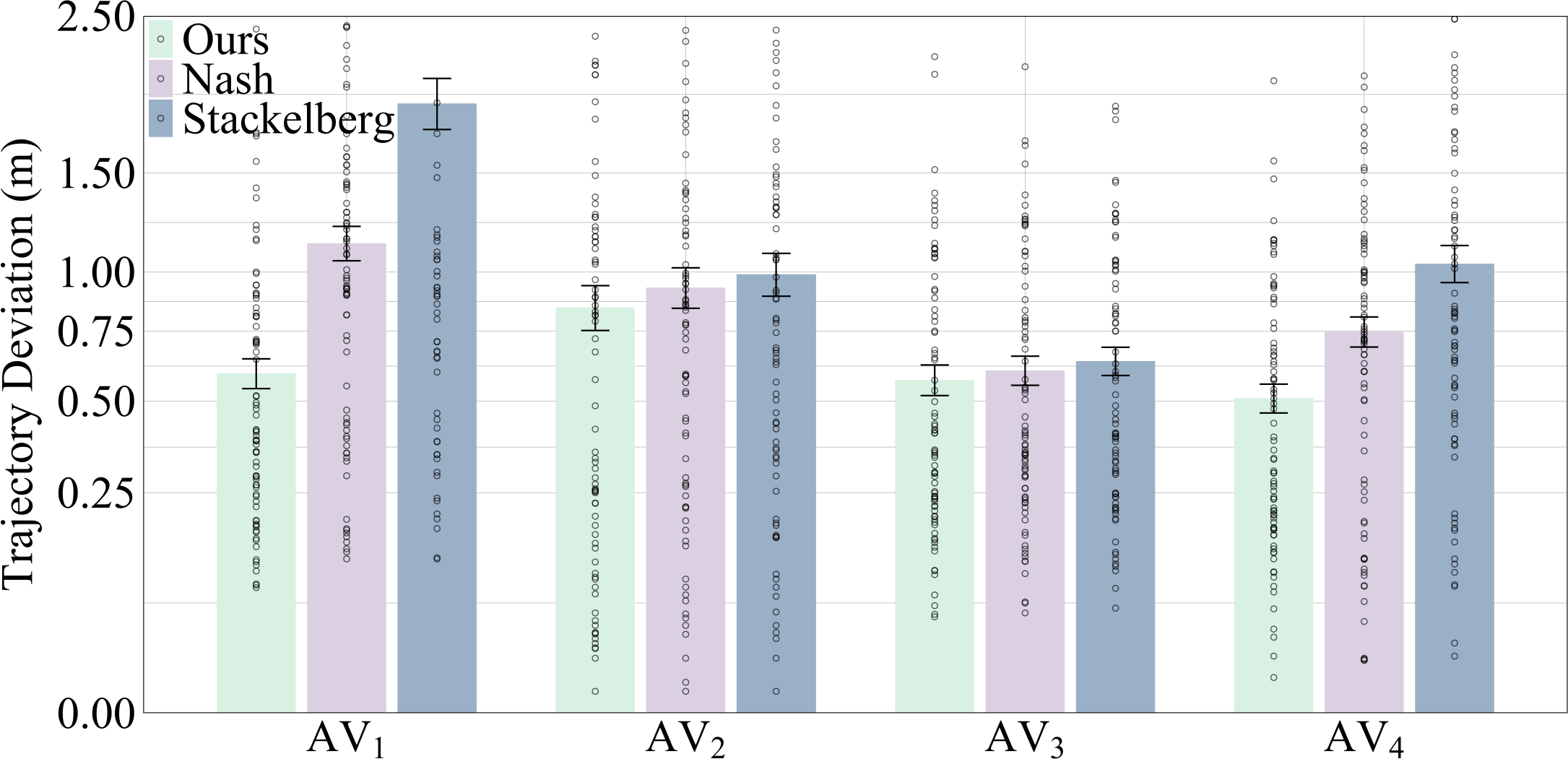}
    \vspace{-2mm}
    \caption{Comparison of trajectory deviations with advanced approaches.}
    \label{fig8_3}
\end{subfigure}%
     \vspace{-2mm}
    \caption{Performance Analysis in Case 2 (ROP = 50\%). }
    \label{fig8}
\end{figure}

\subsection{Case 1: Dual-lane Roundabout (ROP = 100\%)}

The experimental evaluation begins with a baseline scenario featuring a $100\%$ rate of penetration (ROP) at a signal-free, dual-lane roundabout. As depicted in Fig.~\ref{fig6}(a), four CAVs simultaneously approach the roundabout from different directions, creating a complex multi-agent coordination challenge. The conflict zone, highlighted by the red dashed box, marks the critical region where vehicle paths intersect and potential collisions may occur.

Fig.~\ref{fig6}(b) illustrates the velocity profiles of the CAVs under our proposed method. Solid lines indicate the mean velocities, while the shaded bands represent the $95\%$ confidence intervals. The results show that the vehicles maintain stable speeds ranging from $3$ to $5~\mathrm{m/s}$, requiring only minimal speed modulation for safe interaction. This highlights the ability of our MCTS-based framework to ensure both safety and motion smoothness without unnecessary deceleration.

Further insights are provided in Fig.~\ref{fig6}(c), which compares trajectory deviations among our method and two baselines: Nash and Stackelberg. Across all CAVs, our approach consistently yields smaller deviations from the reference trajectories. This significant improvement in tracking accuracy reflects the effectiveness of our planning strategy in preserving intended paths while adhering to safety constraints. Additionally, the reduced size of the error bars indicates greater behavioral consistency and robustness.

\begin{table}[t]
\centering
\setlength{\tabcolsep}{5pt}
\captionsetup{
    labelfont={sc}, 
    textfont={sc}, 
    labelsep=colon, 
    skip=1em,
    singlelinecheck=false,
    justification=centering,
    format=plain
}
\caption{\protect\\ Comparison of Algorithm Performances in Case 1.}
\label{tab2}
\begin{threeparttable}
\resizebox{\linewidth}{!}{
\begin{tabular}{c|c|c|c}
\hline \hline
\multirow{2}{*}{Methods} 
& Average Speed 
& Average Trajectory  
& Collision  \\
& (m/s) 
& Deviation (m)  
& Rate (\%) \\
\hline
Stackelberg 
& $5.22\pm 1.34$  
& $0.89\pm 0.84$    
& $16.0$   \\
\hline
Nash 
& $5.16\pm 1.01$  
& $0.74\pm 0.59$   
& $11.0$   \\
\hline
Baseline  
& $5.44\pm 1.10$  
& $0.65\pm 0.47$  
& $13.0$   \\
\hline
$d_{\max} = 3$ 
& $5.27\pm 1.03$  
& $0.57\pm 0.38$   
& $4.0$   \\
\hline
\rowcolor{lightcyan}
$d_{\max} = 8$ 
& $\mathbf{5.31\pm 0.89}$  
& $\mathbf{0.45\pm 0.39}$   
& $\mathbf{0.0}$   \\
\hline\hline
\end{tabular}}
\end{threeparttable}
\end{table}

The decision-making process is further analyzed through the control input heatmap in Fig.~\ref{fig7}(a), which reveals the temporal evolution of acceleration and yaw rate commands for each vehicle. The predominant green coloring indicates that most control actions are moderate, with occasional stronger interventions (darker colors) occurring primarily during critical in-roundabout driving phases. This pattern demonstrates the framework's ability to generate comfortable trajectories while responding appropriately to dynamic interaction scenarios.

Fig.~\ref{fig7}(b) presents a quantitative assessment of safety performance based on Post-Encroachment Time (PET) distributions. The proposed method with a maximum tree depth of $d_{\text{max}} = 8$ delivers the best outcomes, with no instances of PET falling below the critical threshold of $1.0$ second. In comparison, the baseline approach exhibits $8.8\%$ violations, while the Stackelberg and Nash methods perform worse, with violation rates of $26.1\%$ and $13.2\%$, respectively. These results underscore the effectiveness of our framework, which leverages dynamic risk evaluation and adaptive safety constraints to proactively manage both current and anticipated interactions.

Table~\ref{tab2} demonstrates the superior performance of our proposed method with $d_{\max}=8$, which achieves the lowest collision rate (0.0\%) and the smallest trajectory deviation ($0.45\pm0.39$,m), while maintaining competitive driving speed ($5.31\pm0.89$,m/s). In contrast, Stackelberg and Nash baselines exhibit substantially higher collision rates (16.0\% and 11.0\%), indicating their limited robustness in complex interaction scenarios. Increasing $d_{\max}$ consistently improves safety, confirming the benefit of incorporating larger safety margins in roundabout coordination.

\begin{figure}[t]
    \centering
    \begin{subfigure}{\columnwidth}
    \centering
        \includegraphics[ width=\textwidth]{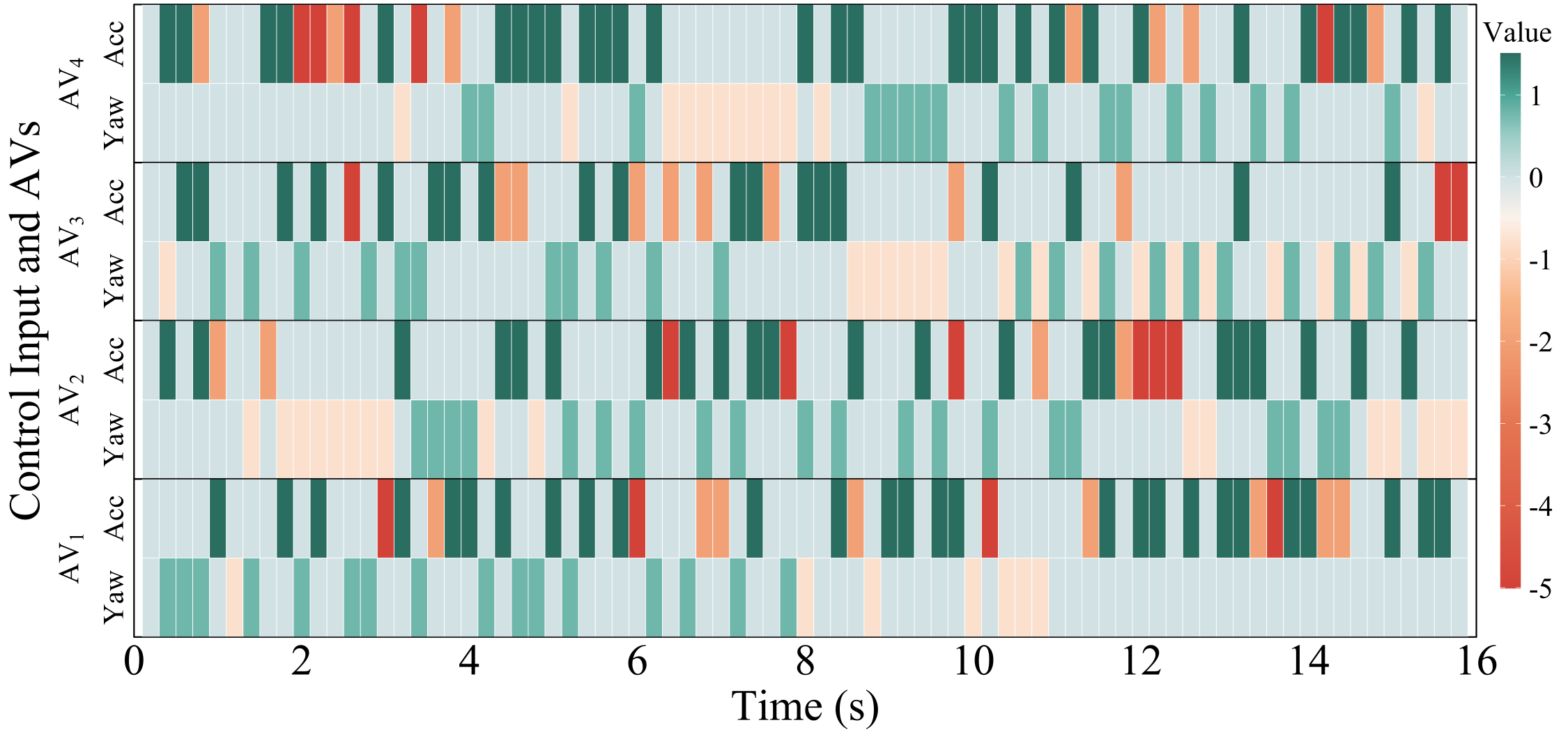}
         \vspace{-6mm}
        \caption{{Heatmap of control inputs for controlled CAVs.}}
        \label{fig9_1}
    \end{subfigure}
 
\begin{subfigure}{\columnwidth}
\centering
    \includegraphics[width=1\textwidth]{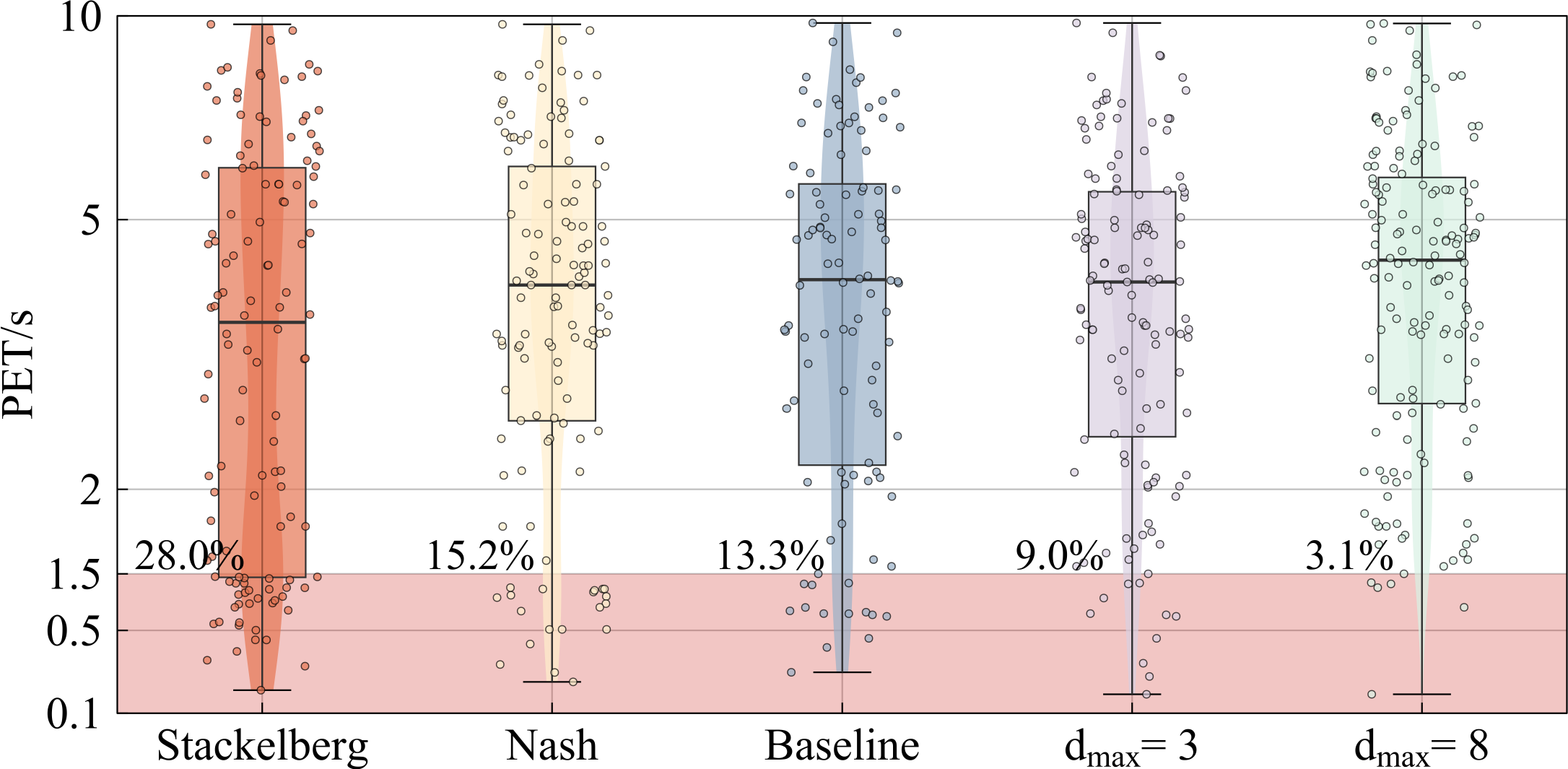}
    \vspace{-5mm}
    \caption{PET comparison.}
    \label{fig9_2}
\end{subfigure}%
     \vspace{-2mm}
    \caption{Analysis of decision-making and safety performance. (a) Variations in control inputs. (b) PET distributions and violations: benchmarks, Baseline (without adaptive risk evaluation), and our methods with $d_{\text{max}}$ of 3 and 8 (ROP = 50\%). }
    \label{fig9}
\end{figure}

\begin{table}[t]
\centering
\setlength{\tabcolsep}{5pt}
\captionsetup{
    labelfont={sc}, 
    textfont={sc}, 
    labelsep=colon, 
    skip=1em,
    singlelinecheck=false,
    justification=centering,
    format=plain
}
\caption{\protect\\ Comparison of Algorithm Performances in Case 2.}
\label{tab3}
\begin{threeparttable}
\resizebox{\linewidth}{!}{
\begin{tabular}{c|c|c|c}
\hline \hline
\multirow{2}{*}{Methods} 
& Average Speed 
& Average Trajectory  
& Collision  \\
& (m/s) 
& Deviation (m)  
& Rate (\%) \\
\hline
Stackelberg 
& $5.25\pm 1.31$  
& $1.14\pm 1.06$    
& $33.0$   \\
\hline
Nash 
& $5.12\pm 1.24$  
& $0.85\pm 0.68$   
& $29$   \\
\hline
Baseline  
& $4.83\pm 0.82$  
& $0.81\pm 0.65$  
& $17$   \\
\hline
$d_{\max} = 3$ 
& $5.16\pm 0.95$  
& $0.76\pm 0.62$   
& $5.0$   \\
\hline
\rowcolor{lightcyan}
$d_{\max} = 8$ 
& $\mathbf{5.23\pm 0.94}$  
& $\mathbf{0.63\pm 0.59}$   
& $\mathbf{2.0}$   \\
\hline\hline
\end{tabular}}
\end{threeparttable}
\end{table}

\subsection{Case 2: Dual-lane Roundabout (ROP = 50\%)}

To assess the robustness of our proposed framework under mixed traffic conditions, we conduct experiments with a $50\%$ CAV penetration rate, where CAVs and HDVs coexist and interact at the signal-free, dual-lane Roundabout. As illustrated in Fig.~\ref{fig8}(a), the test setup features four CAVs and four HDVs entering the roundabout from different entrance ports, introducing additional coordination complexity due to the unpredictable behavior of human drivers.

The velocity profiles depicted in Fig.~\ref{fig8}(b) highlight the effectiveness of our method in managing such complex roundabout. In contrast to the full-CAV scenario, the velocity trajectories exhibit greater variability—ranging between $3$--$5~\mathrm{m/s}$—and wider confidence intervals, reflecting the influence of HDV-induced uncertainty. Despite this, the velocity transitions remain relatively smooth, suggesting that our MCTS-based framework is capable of adapting to human behaviors while preserving both safety and traffic flow efficiency.

As shown in Fig.~\ref{fig8}(c), the trajectory deviation analysis further confirms the advantages of our approach over baseline methods. Our framework consistently yields smaller deviations from reference paths, demonstrating improved path-tracking accuracy even in the presence of mixed vehicle types. These results underscore the effectiveness of our risk-aware decision-making mechanism and its adaptability in complex, uncertain traffic environments.

The control input heatmap in Fig.~\ref{fig9}(a) reveals a greater diversity and frequency of CAV behavior adjustments compared to the fully autonomous scenario. The emergence of more pronounced light and dark regions indicates that CAVs engage in more dynamic control actions in response to the unpredictability of human-driven vehicles. This pattern reflects the adaptive nature of our framework, which effectively modulates between assertive and conservative behaviors to ensure safe and efficient interaction with HDVs.

\begin{figure}[t]
    \centering
    \includegraphics[width=1\linewidth]{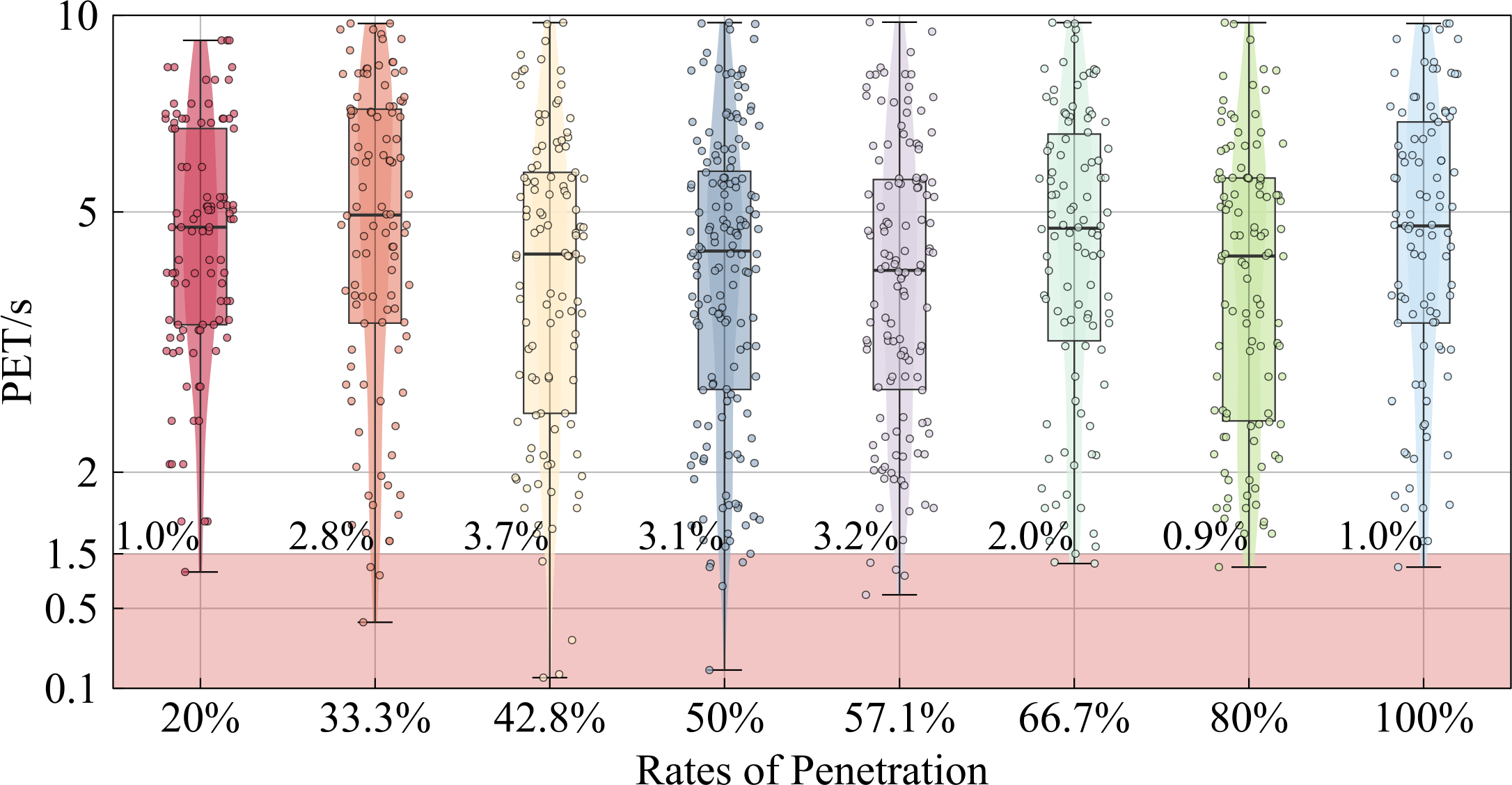}
    \vspace{-5mm}
    
\caption{PET violations under different CAV penetration rates.}
    \label{pte_all}
\end{figure}

The safety performance assessment based on PET distributions, as illustrated in Fig.~\ref{fig9}(b), underscores the difficulties posed by mixed traffic scenarios. Our method with $d_{\text{max}} = 8$ achieves the most favorable safety outcome, with only $3.1\%$ of instances violating the PET threshold. In contrast, the baseline method results in a notably higher violation rate of $13.3\%$, while the Stackelberg and Nash strategies perform worse still, with violation rates of $28.0\%$ and $15.2\%$, respectively. These findings emphasize the complexity of maintaining safe interactions when human-driven vehicles are involved.

As shown in Table~\ref{tab3}, our MCTS-based method with $d_{\max}=8$ achieves the best overall performance in Case~2, yielding the lowest collision rate (2.0\%) and the smallest trajectory deviation ($0.63\pm0.59$,m), while maintaining a high average speed ($5.23\pm0.94$,m/s). In comparison, Stackelberg and Nash exhibit substantially higher collision rates (33.0\% and 29.0\%), highlighting their limited robustness in mixed traffic with human-driven uncertainty. These results confirm that increasing $d_{\max}$ significantly improves safety and motion consistency in complex roundabout interactions.

To evaluate safety under varying CAV penetration rates, experiments were conducted with penetration rates ranging from $20\%$ to $100\%$. Fig.~\ref{pte_all} shows the PET distributions, revealing improved safety as CAV penetration increases. At low penetration rates ($20\%$-$33.3\%$), PET distributions are more variable, with low violation rates ($1.0\%$-$3.7\%$) due to the unpredictability of HDVs. In the medium range ($42.8\%$–$57.1\%$), violation rates initially rise to $3.7\%$ at $42.8\%$, then stabilize around $3.2\%$, reflecting the complexity of mixed traffic. At high penetration rates ($66.7\%$-$100\%$), violation rates drop to $0.9\%$, and PET distributions narrow, demonstrating consistent safety margins due to the dominance of CAV behavior.

\begin{figure}[t]
\centering

\begin{subfigure}[b]{0.48\linewidth}
    \centering
    \includegraphics[width=\linewidth]{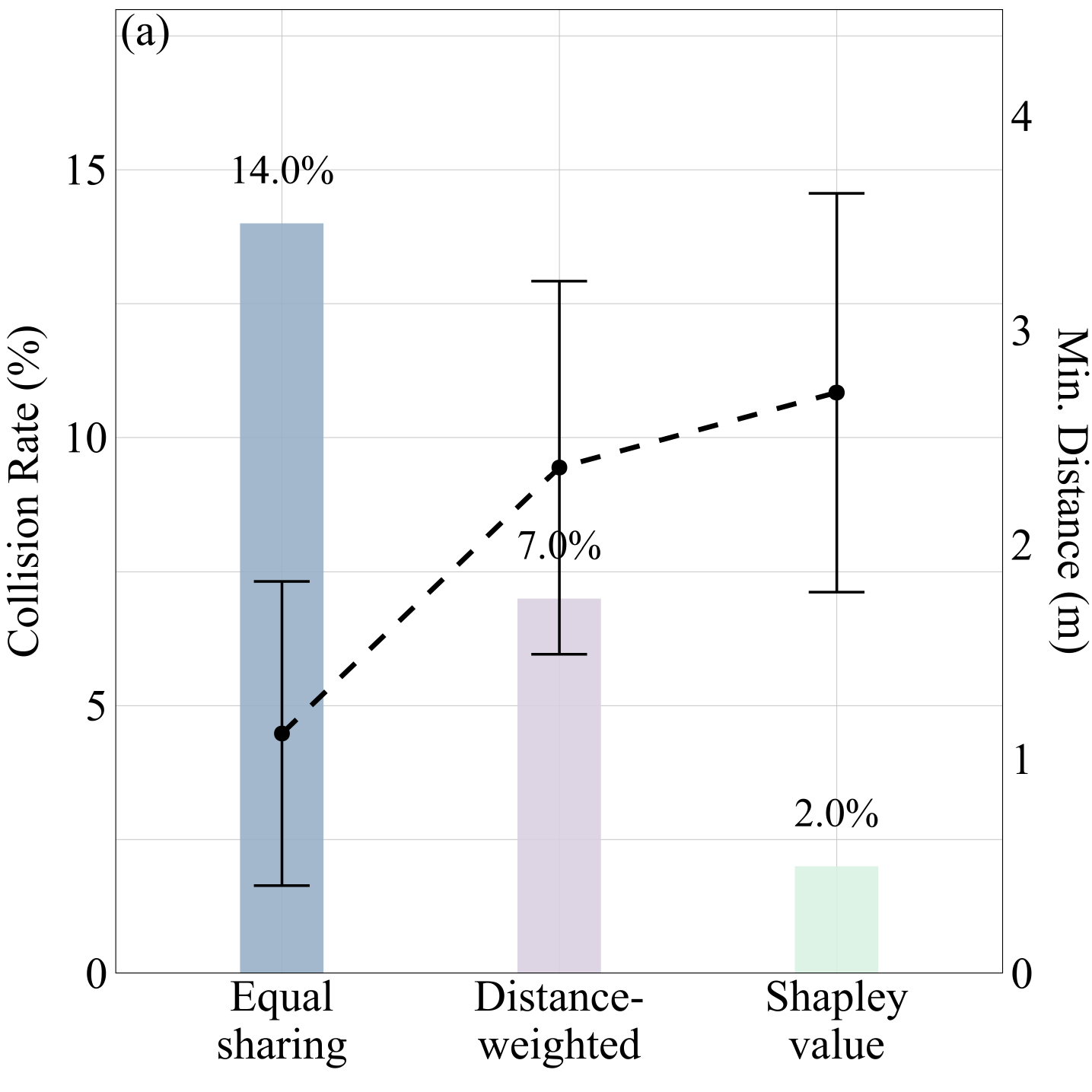}
\end{subfigure}
\hfill
\begin{subfigure}[b]{0.48\linewidth}
    \centering
    \includegraphics[width=\linewidth]{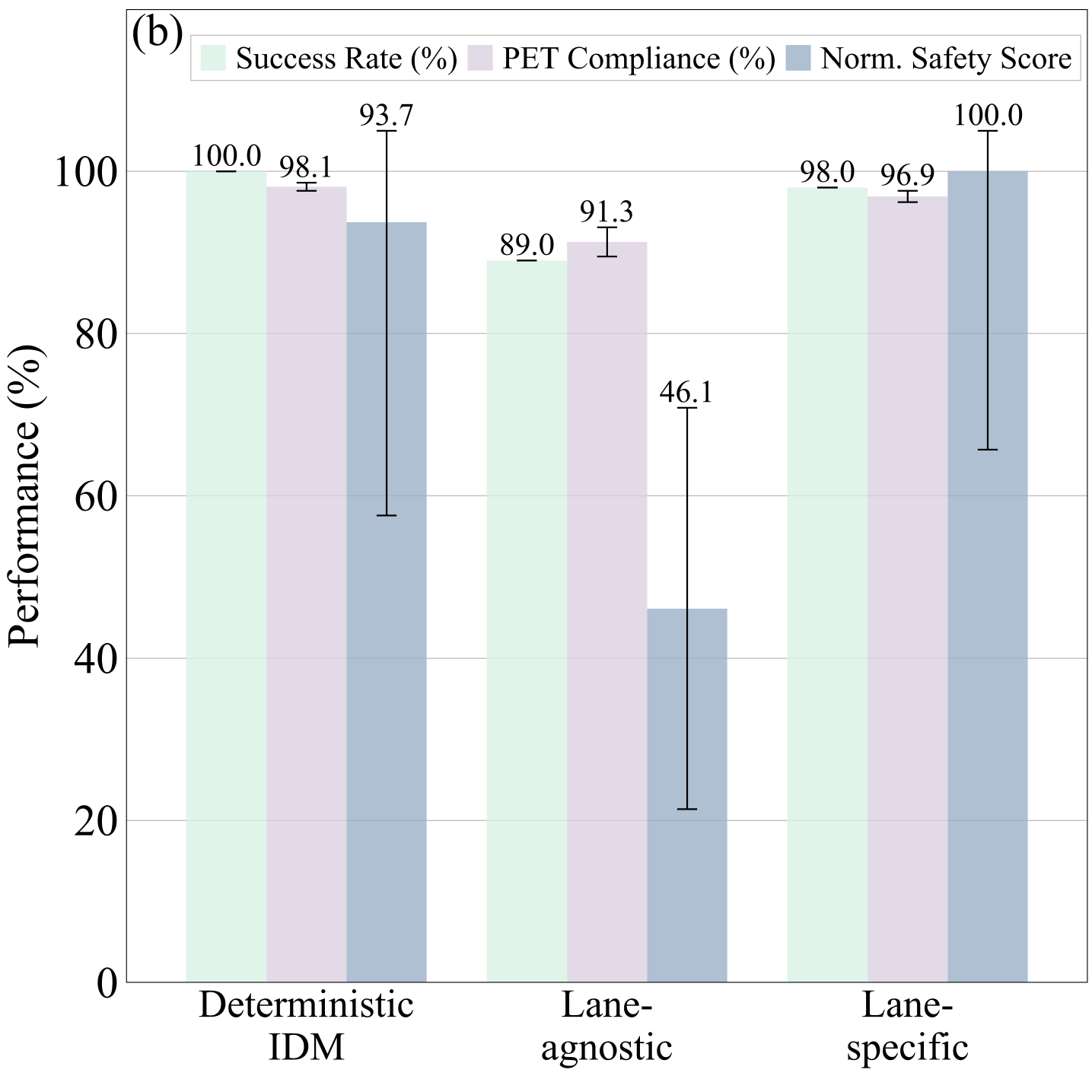}
\end{subfigure}

\vspace{-2mm}
\caption{Ablation results. (a) Reward allocation strategies. (b) HDV uncertainty models.}
\label{fig12_ablation_all}
\end{figure}

\begin{table}[t]
\centering
\captionsetup{
    labelfont={sc}, 
    textfont={sc}, 
    labelsep=colon, 
    singlelinecheck=false,
    justification=centering,
    format=plain
}
\caption{Sensitivity and ablation analysis}
\label{tab:sensitivity_ablation}
\setlength{\tabcolsep}{2pt}
\renewcommand{\arraystretch}{1.0}
\small
\begin{tabular}{lcccc}
\hline
\hline
\multirow{2}{*}{Setting} 
& Collision 
& PET Viol. 
& Traj. Dev.
& Min. \\
& Rate (\%) 
& Rate (\%)
& (m) 
& Distance (m) \\
\midrule
\multicolumn{5}{l}{{\footnotesize \textit{Min. Safe Distance $d_{\min}$ (Default $d_{\min} = 2.0$\,m)}}} \\
$d_{\min} = 1.0$\,m & 9.0 & \spm{8.2}{0.9} & \spm{0.56}{0.39} & \spm{1.31}{0.74} \\
\rowcolor{lightcyan}
$\bm{d_{\min} = 2.0}$\,\textbf{m}
& \textbf{2.0}
& \textbf{\spm{3.1}{0.7}}
& \textbf{\spm{0.73}{0.61}}
& \textbf{\spm{2.72}{0.93}} \\
$d_{\min} = 3.0$\,m & 1.0 & \spm{2.4}{0.5} & \spm{1.42}{0.85} & \spm{3.61}{1.12} \\
\midrule
\multicolumn{5}{l}{{\footnotesize \textit{Reward Allocation Strategy (Default Shapley)}}} \\
Equal sharing            & 14.0 & \spm{10.2}{2.1} & \spm{1.63}{0.85} & \spm{1.12}{0.71} \\
Distance-weighted        & 7.0 & \spm{5.3}{1.6} & \spm{1.12}{0.78} & \spm{2.36}{0.87} \\
\rowcolor{lightcyan}
\textbf{Shapley value}
& \textbf{2.0}
& \textbf{\spm{3.1}{0.7}}
& \textbf{\spm{0.73}{0.61}}
& \textbf{\spm{2.71}{0.93}} \\
\midrule
\multicolumn{5}{l}{{\footnotesize \textit{HDV Uncertainty Model (Default Lane-Specific)}}} \\
Deterministic IDM                          & 0.0 & \spm{1.9}{0.5} & \spm{0.52}{0.43} & \spm{2.54}{0.98} \\
Lane-agnostic & 11.0 & \spm{8.7}{1.8} & \spm{1.66}{0.82} & \spm{1.25}{0.67} \\
\rowcolor{lightcyan}
\textbf{Lane-specific}
& \textbf{2.0}
& \textbf{\spm{3.1}{0.7}}
& \textbf{\spm{0.73}{0.61}}
& \textbf{\spm{2.71}{0.93}} \\
\hline
\hline
\end{tabular}
\end{table}


\subsection{Sensitivity and Ablation Analysis}
\label{sec:sensitivity_ablation}

To evaluate the robustness of key design choices and validate the
contributions of individual components, we conduct sensitivity and
ablation studies. The quantitative results are summarized in
Table~\ref{tab:sensitivity_ablation}, with visual comparisons for the
two core algorithmic contributions presented in
Fig.~\ref{fig12_ablation_all}.

\subsubsection{Sensitivity to Minimum Safe Distance}
We vary the minimum safe distance threshold
$d_{\min} \in \{1.0, 2.0, 3.0\}$\,m. As shown in
Table~\ref{tab:sensitivity_ablation}, a smaller threshold
($d_{\min} = 1.0$\,m) permits closer interactions, resulting in
elevated collision rates ($9.0\%$) and reduced minimum distances
($1.31$\,m). Conversely, a larger threshold ($d_{\min} = 3.0$\,m)
yields slightly lower collision rates ($1.0\%$) but introduces
excessive conservatism, increasing trajectory deviations to
$1.42$\,m. The default setting $d_{\min} = 2.0$\,m achieves an
effective balance between safety and efficiency, maintaining a low
collision rate ($2.0\%$) with moderate trajectory deviations
($0.73$\,m).

\subsubsection{Reward Allocation Strategy}
We compare three reward allocation strategies: equal sharing,
distance-weighted sharing, and Shapley value allocation. As
illustrated in Fig.~\ref{fig12_ablation_all}(a), the Shapley value
method significantly outperforms both alternatives. Equal sharing,
which ignores individual contributions, leads to the highest
collision rate ($14.0\%$) and PET violation rate ($10.2\%$), as
agents lack incentive for cooperative behavior. Distance-weighted
sharing improves upon equal sharing but still yields a $7.0\%$
collision rate. The Shapley value method reduces collision rates to
$2.0\%$ and achieves the best minimum distance ($2.71$\,m),
demonstrating that fairly attributing each agent's marginal
contribution to the coalition reward is essential for effective
multi-agent coordination.

\subsubsection{HDV Uncertainty Model}
We evaluate three uncertainty modeling approaches for human-driven
vehicles: deterministic IDM, lane-agnostic uncertainty, and
lane-specific uncertainty. As shown in
Fig.~\ref{fig12_ablation_all}(b), the deterministic IDM achieves
zero collisions in simulation, which is expected since the
prediction model matches the simulator. However, this zero-margin
approach offers no robustness against real-world behavioral
deviations. The lane-agnostic model, which applies uniform
uncertainty across lanes, introduces excessive conservatism that
paradoxically increases collisions to $11.0\%$, as overly cautious
planning causes hesitation and deadlock. Our lane-specific model
strikes an effective balance, reducing collisions to $2.0\%$ while
providing structured robustness margins calibrated to lane-dependent
behavioral variability, as outer-lane vehicles exhibit inherently
higher trajectory uncertainty than inner-lane vehicles.

\section{Conclusion}
\label{sec5}
This paper introduces a safety-critical decision-making framework 
for autonomous vehicles navigating unsignalized, dual-lane 
roundabouts by integrating Monte Carlo Tree Search (MCTS) with a 
hierarchical risk assessment strategy. The framework offers three 
major innovations: a multi-agent MCTS structure for scalable and 
efficient action space exploration, a hierarchical safety assessment 
mechanism for robust spatiotemporal risk evaluation, and an adaptive 
reward function that effectively balances safety and efficiency. 
Experimental results confirm the effectiveness of the proposed method 
under varying autonomous vehicle penetration rates. In fully 
autonomous settings ($100\%$ CAVs), the framework achieves reduced 
trajectory deviations and eliminates PET violations when compared to 
baseline approaches. In mixed traffic scenarios ($50\%$ CAVs + $50\%$ 
HDVs), the framework delivers even greater improvements by reliably 
handling the uncertainty introduced by human drivers, while 
maintaining low deviation and high safety margins. 
Sensitivity and ablation analyses further demonstrate 
robustness to key parameter choices and confirm that each proposed 
component---Shapley value allocation and lane-specific uncertainty 
modeling---contributes meaningfully to overall performance.
We note that the current framework assumes accurate state observations 
and plans independently per CAV without inter-vehicle communication. 
While these assumptions are idealized, the probabilistic C2H 
uncertainty model provides inherent robustness margins that can 
partially absorb sensing inaccuracies. The current results, validated 
in simulation, demonstrate the framework's promise as an effective 
decision-support tool for complex roundabout coordination.
Future work will focus on improving computational scalability 
through parallelized search strategies, extending the framework to 
diverse roundabout geometries, and integrating 
data-driven behavior prediction with validation under realistic 
sensing noise and communication delays.

\bibliographystyle{IEEEtran_custom}
\bibliography{zq_lib}

\begin{IEEEbiography}[{\includegraphics[width=1in,height=1.25in,clip,keepaspectratio]{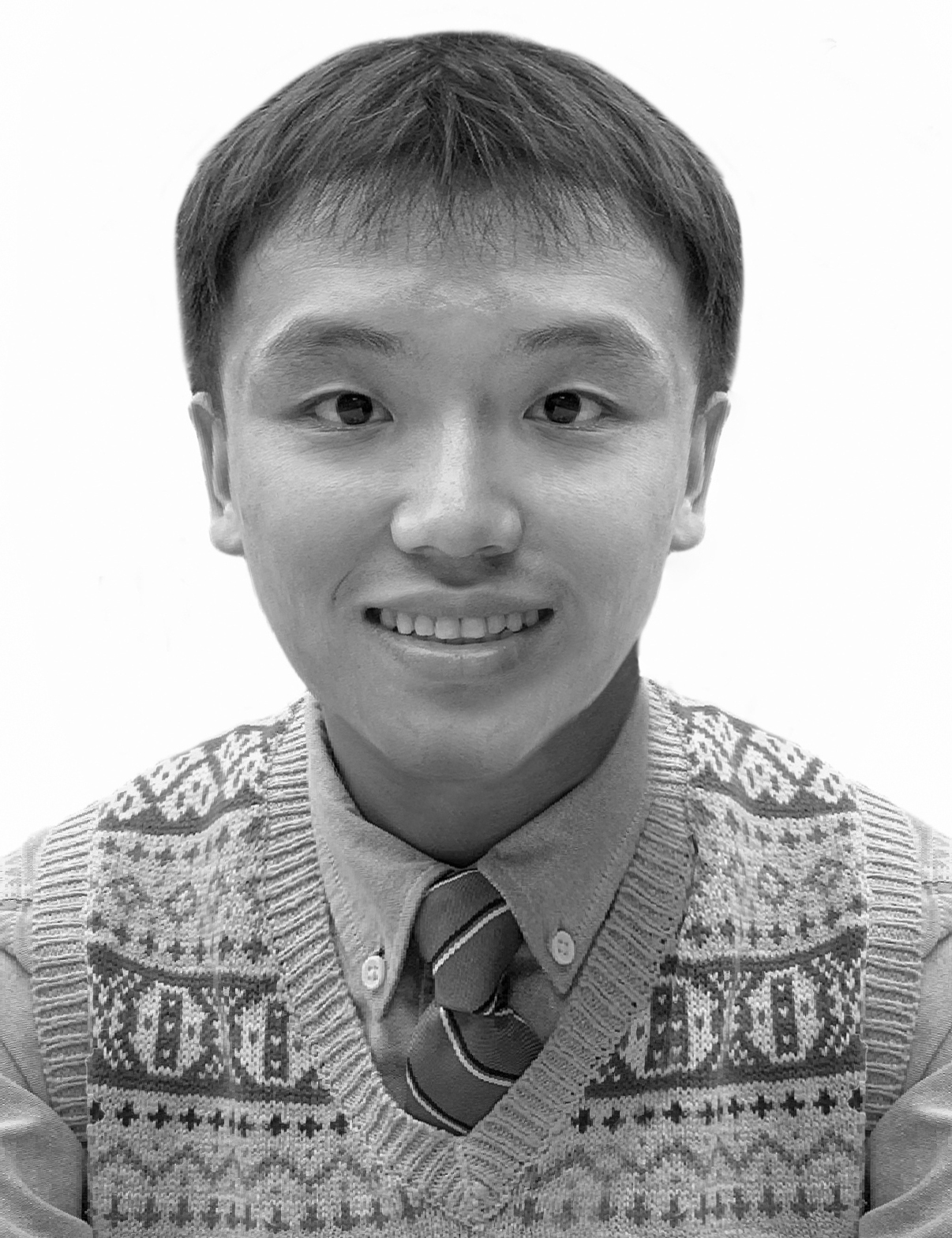}}]%
{Zhihao Lin}
received the M.S. degree from the College of Electronic Science and Engineering, Jilin University, Changchun, China, in 2019. He is currently pursuing the Ph.D. degree with the James Watt School of Engineering, University of Glasgow, UK. His research interests focus on Structure-aware Reinforcement Learning and Multi-Agent Decision Making.
\end{IEEEbiography}
\vspace{-5 mm}

\vspace{-5 mm}
\begin{IEEEbiography}
[{\includegraphics[width=1in,height=1.25in, clip]{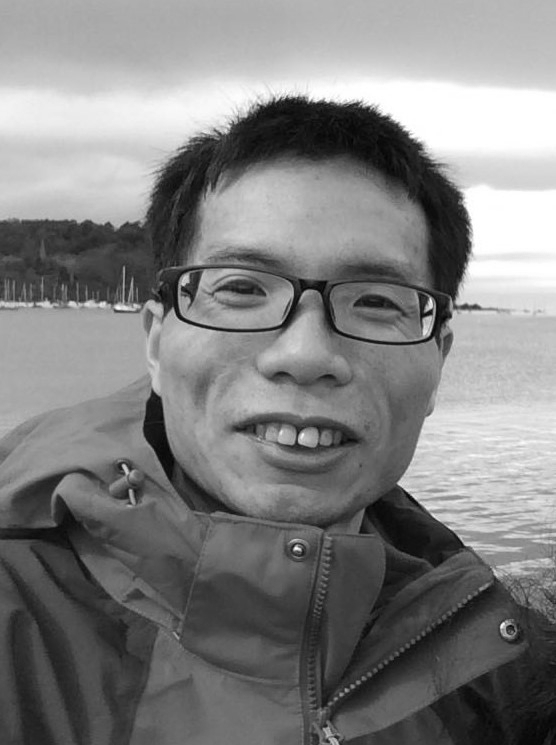}}]{Jianglin Lan} received the Ph.D. degree from the University of Hull in 2017. He has been a Leverhulme Early Career Fellow and Lecturer at the University of Glasgow since 2022. He was a Visiting Professor at the Robotics Institute, Carnegie Mellon University, in 2023. From 2017 to 2022, he held postdoc positions at Imperial College London, Loughborough University, and University of Sheffield. His research interests include AI, optimisation, control theory, and autonomy.
\end{IEEEbiography}

\vspace{-5 mm}
\begin{IEEEbiography}[{\includegraphics[width=1in,height=1.25in,clip,keepaspectratio]{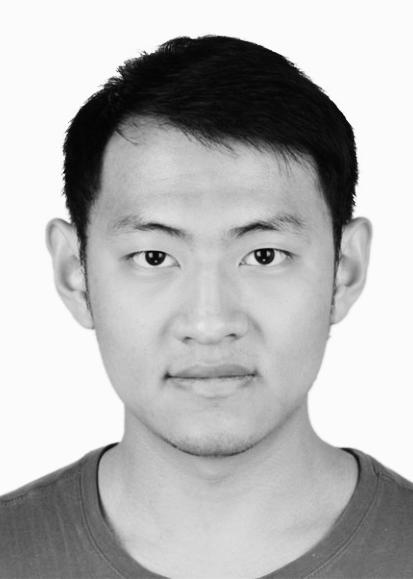}}]%
{Shuo Liu}
(Student Member, IEEE) received his M.S. degree in Mechanical Engineering from Columbia University, New York, NY, USA, in 2020 and his B.Eng. degree in Mechanical Engineering from Chongqing University, Chongqing, China, in 2018. He is currently a Ph.D. candidate in Mechanical Engineering at Boston University, Boston, USA. His research interests include optimization, nonlinear control, deep learning, and robotics.
\end{IEEEbiography}

\vspace{-5 mm}

\begin{IEEEbiography}[{\includegraphics[width=1in,height=1.25in,clip]{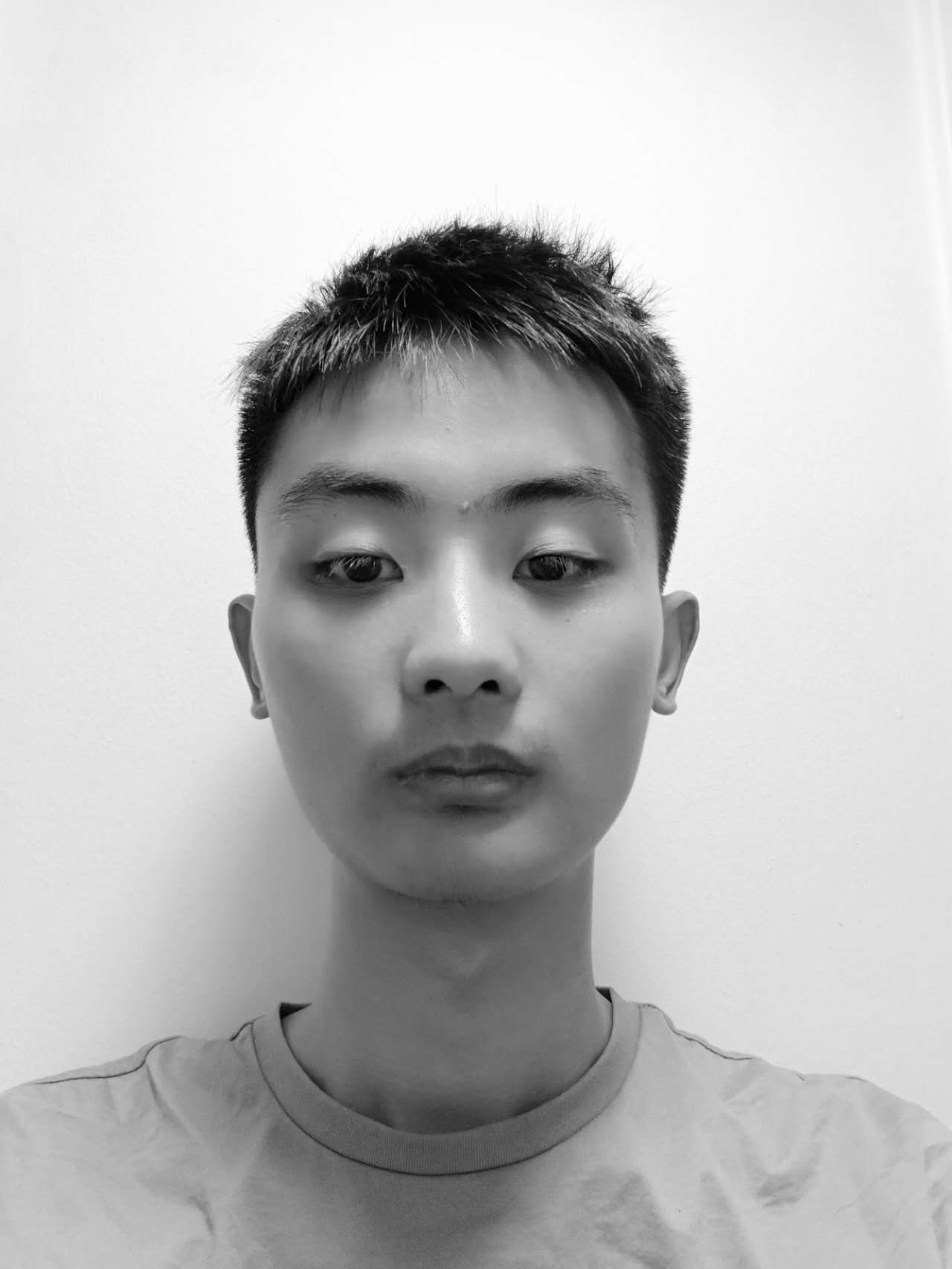}}]{Zhen Tian} received the B.Eng. degree in Electronic and Electrical Engineering from the University of Strathclyde, Glasgow, U.K., in 2020, and the Ph.D. degree from the College of Science and Engineering, University of Glasgow, Glasgow, U.K., in 2025. He is currently a postdoctoral researcher at the University of Glasgow. His research interests include interactive vehicle decision systems and autonomous racing strategies.
\end{IEEEbiography}

\vspace{-5 mm}

\begin{IEEEbiography}[{\includegraphics[width=1in,height=1.25in,clip,keepaspectratio]{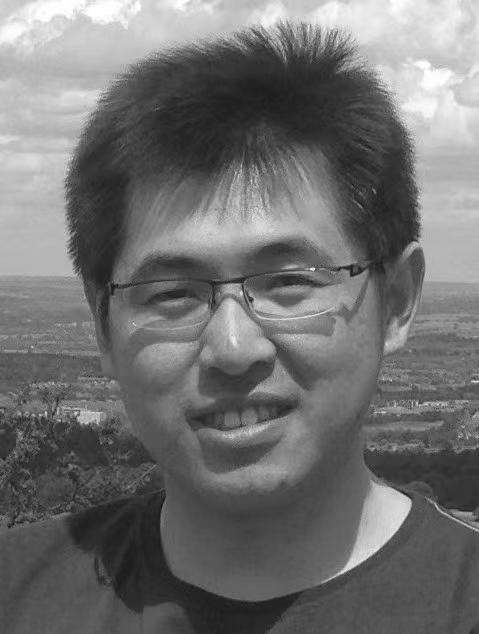}}]%
{Dezong Zhao}
received the B.Eng. and M.S. degrees
from Shandong University, Jinan, China, in 2003
and 2006, respectively, and the Ph.D. degree from Tsinghua University, Beijing, China, in 2010, all in Control Science and Engineering. He is a Reader in Autonomous Systems with the James Watt School of Engineering, University of Glasgow and a Turing Fellow with the Alan Turing Institute. He was awarded a Royal Society-Newton Advanced Fellow in 2020 and an EPSRC Innovation Fellow in 2018.
\end{IEEEbiography}

\vspace{-5 mm}
\begin{IEEEbiography}[{\includegraphics[width=1in,height=1.25in,clip,keepaspectratio]{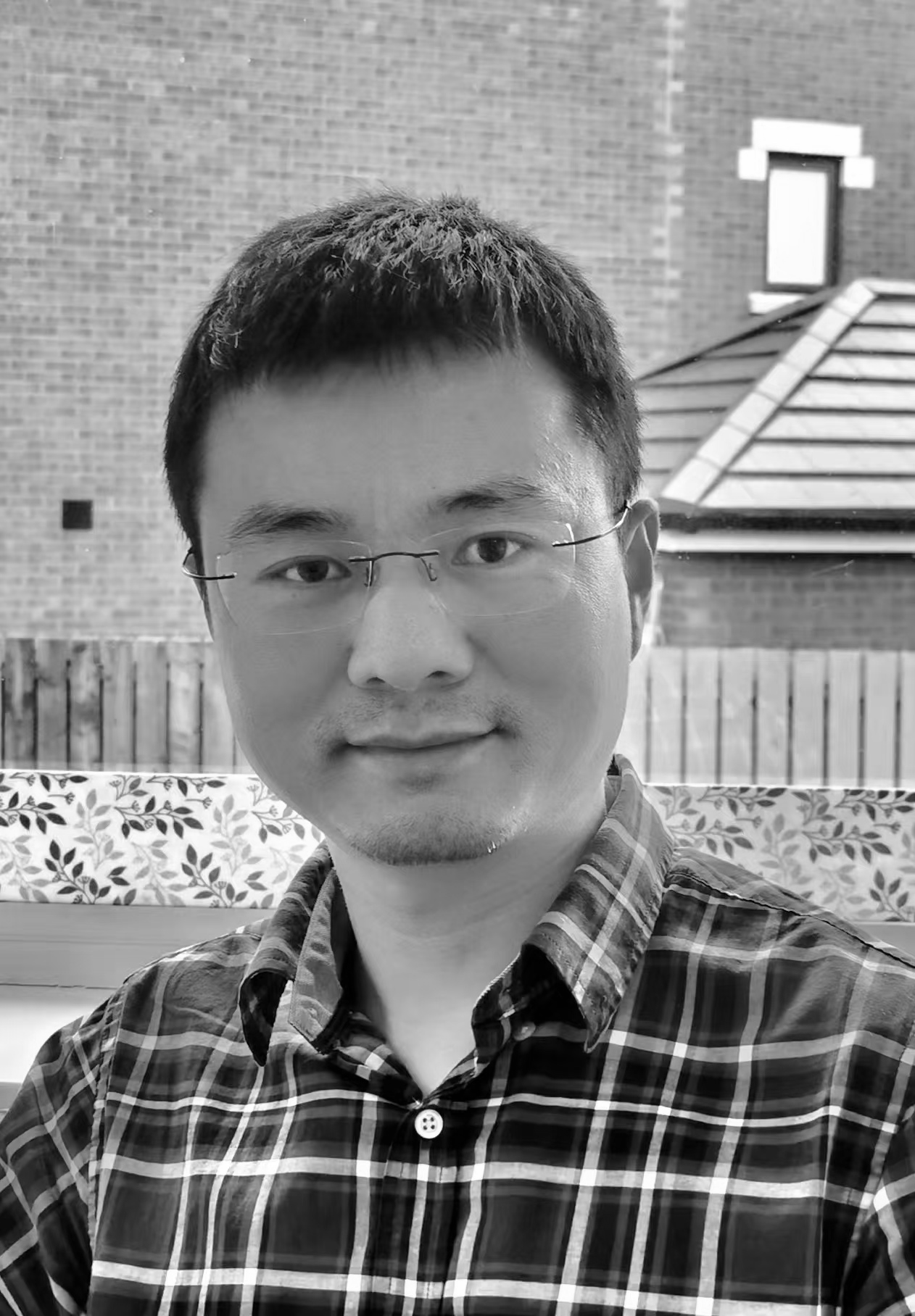}}]%
{Chongfeng Wei} received his Ph.D. degree in mechanical engineering from the University of Birmingham in 2015. He is now an Associate Professor (University Senior Lecturer) at University of Glasgow, UK. His current research interests include decision-making and control of intelligent vehicles, human-centric autonomous driving, cooperative automation, and dynamics and control of mechanical systems. He is also serving as an Associate Editor of IEEE TITS, IEEE TIV, IEEE TVT, and Frontier on Robotics and AI.
\end{IEEEbiography}

\end{document}